\documentclass[%
 reprint,
 superscriptaddress,
 numerical,
 showpacs,
 amsmath,amssymb,
 aps,
 longbibliography,
 prb,
 floatfix
]{revtex4-2}

\usepackage{graphicx}
\usepackage{xcolor}

\usepackage[normalem]{ulem}

\usepackage[%
  colorlinks=true,
  urlcolor=blue,
  linkcolor=blue,
  citecolor=blue
]{hyperref}

\let\Re\relax
\let\Im\relax

\DeclareMathOperator{\diag}{diag}
\DeclareMathOperator{\tr}{tr}

\DeclareMathOperator{\Re}{Re}
\DeclareMathOperator{\Im}{Im}

\newcommand{\ket}[1]{| #1 \rangle}
\newcommand{\bra}[1]{\langle#1 |}
\newcommand{\braket}[2]{\langle#1 | #2 \rangle}

\begin{document}
\title{Berry connection induced anomalous wave-packet dynamics in non-Hermitian systems}

\date{\today}

\author{Navot Silberstein}
\affiliation{Raymond and Beverly Sackler School of Physics and Astronomy,
Tel-Aviv University, Tel-Aviv 6997801, Israel}

\author{Jan Behrends}
\affiliation{T.C.M. Group, Cavendish Laboratory, University of Cambridge, J.J. Thomson Avenue, Cambridge, CB3 0HE, United Kingdom}
\author{Moshe Goldstein}
\affiliation{Raymond and Beverly Sackler School of Physics and Astronomy,
Tel-Aviv University, Tel-Aviv 6997801, Israel}

\author{Roni Ilan}
\affiliation{Raymond and Beverly Sackler School of Physics and Astronomy,
Tel-Aviv University, Tel-Aviv 6997801, Israel}

\begin{abstract}
Berry phases strongly affect the properties of crystalline materials, giving rise to modifications of the semiclassical equations of motion that govern wave-packet dynamics.
In non-Hermitian systems, generalizations of the Berry connection have been analyzed to characterize the topology of these systems.
While the topological classification of non-Hermitian systems is being developed, little attention has been paid to the impact of the new geometric phases on dynamics and transport.
In this work, we derive the full set of semiclassical equations of motion for wave-packet dynamics in a system governed by a non-Hermitian Hamiltonian, including corrections induced by the Berry connection.
We show that non-Hermiticity is manifested in anomalous weight rate and velocity terms that are present already in one-dimensional systems, in marked distinction from the Hermitian case.
We express the anomalous weight and velocity in terms of the Berry connections defined in the space of left and right eigenstates and compare the analytical results with numerical lattice simulations.
Our work specifies the conditions for observing the anomalous contributions to the semiclassical dynamics and thereby paves the way to their experimental detection, which should be within immediate reach in currently available metamaterials.
\end{abstract}

\maketitle

\section{Introduction}

Topological band theory is an extension of the conventional band theory of solids that accounts for possible nontrivial topology of the band structure of noninteracting Hamiltonians describing particles under the influence of periodic potentials~\cite{Hasan:2010ku,Qi:2011hb,Bansil:2016bu}.
A central quantity that emerges is the geometric phase, or the Berry phase, that is accumulated by the wave functions via spectral flow along closed trajectories in momentum space~\cite{Berry:1984ka,Zak:1989fh}. The corresponding Berry curvature, which functions as an analog of a magnetic field in reciprocal space, endows isolated bands with a topological index, the Chern number, that is forced to be quantized when integrated over the entire Brillouin zone~\cite{Thouless:1982kq,Haldane:1988gh}.

Being able to consider the Berry curvature of a single isolated band strongly relies on the adiabatic theorem~\cite{Born:1928hd,Xiao:2010kw}. For a single band to have a well-defined and nondivergent Berry curvature, the band has to be isolated from other bands such that adiabatic evolution or a spectral flow can be carried out confined within a single band, i.e., slower than the time scale set by the smallest gap to other bands ~\cite{Xiao:2010kw,Thouless:1983hb}. The single-band Berry curvature is also an essential ingredient in characterizing transport. The semiclassical equations of motion describing wave-packet dynamics are augmented by an anomalous velocity term that is directly proportional to it~\cite{Chang1995,Chang1996,Sundaram:1999ht}. This is the underlying principle of phenomena such as the intrinsic contribution to the anomalous Hall effect in systems with broken time reversal symmetry~\cite{Jungwirth:2002fk,Nagaosa:2010js}.

Since certain systems subjected to dissipation or energy gain can be described by a non-Hermitian Hamiltonian~\cite{Bender:2007kr,Rotter:2009fr,Miri:2019cf,ElGanainy:2019ie,Ozawa:2019ij}, the topological classification of such Hamiltonians may answer questions related to transport and boundary states~\cite{Bernard2002,Magnea:2008gd,Shen:2018bx,Gong:2018ko,MartinezAlvarev:2018kg,Kawabata:2019en,Kawabata:2019fd,Kunst:2019jz,Ghatak:2019cs,Lieu:2020bc}.
A natural extension of the standard definitions, however, should be done with caution, as the adiabatic theorem for non-Hermitian Hamiltonians may break down~\cite{Kvitsinsky:1991km,Nenciu:1992cx,Uzdin:2011dk,Berry:2011ba,Ibanez:2014fv,Milburn:2015hx} (as we highlight in the following and describe in detail in Appendix~\ref{sec:adiabatic_failure}). Moreover, the Hamiltonian possesses in general different right and left eigenvectors, allowing for several distinct generalizations of the Berry connection~\cite{Garrison:1988dt,Dattoli:1990cp,Sun:1993ip,Shen:2018bx}.
And indeed, standard rules such as the bulk-boundary correspondence enforcing the existence of edge states for systems with Chern bands become complicated~\cite{Kunst:2018ku,Herviou:2019ih,Jin:2019iy,Zhang:2019ej,Borgnia:2020hi}, and sometimes do not survive with non-Hermiticity~\cite{Xiong:2018ch,Yao:2018cj,MartinezAlvarez:2018fh}.

The fallacies of common conceptions and intuitions regrading topology and its characteristics in non-Hermitian systems raise profound questions:
When the adiabatic theorem seems to fail, how should one treat wave-packet dynamics in Bloch bands?
In particular, when the system is governed by a non-Hermitian Hamiltonian, is the single-band Berry curvature a meaningful quantity to consider, and under which conditions can one still disentangle the bands and isolate the effects it contributes to? Does the appearance of a generalized anomalous velocity in the equations of motion yield novel observable effects unparalleled in Hermitian systems? Are there additional anomalous contributions unique to this setting, and if so, what are their consequences?

In this work, we set to address the above questions. We derive the generalized semiclassical equations of motion for a wave packet whose dynamics is governed by a non-Hermitian Hamiltonian, including all terms induced by the Berry connection. We analyze the evolution of the wave packet in the presence of electric fields and show that, in stark contrast with such dynamics in Hermitian systems, the dynamics in non-Hermitian systems is characterized by three equations of motion describing the evolution of the average momentum, velocity and weight rate of the wave packet, the latter two supplemented by anomalous terms. These terms are expressed as combinations of the generalized Berry connections that are gauge invariant and vanish or reduce to previously known terms when the system is Hermitian. The same terms may be linked to first-order corrections in perturbation theory to the velocity~\cite{Sternheim:1972ix}, similar to the anomalous velocity in two- and higher-dimensional Hermitian systems~\cite{Karplus:1954bv,Kohn:1957ev}. We carefully examine the role of single-band quantities and show that the anomalous terms we find contribute to wave-packet dynamics of the band with the largest imaginary part of the energies. We thereby establish the conditions under which the equations of motions are valid in their single-band form. As we argue, in periodic media there are stringent limitations for using single-band quantities over the entire Brillouin zone, related to crossings of the imaginary part of the energies and accordingly to the failure of the adiabatic theorem. These limitations, as well as the full set of equations of motion will be manifested in future studies of transport in non-Hermitian systems, which is currently only poorly understood.

As a first testable consequence of the dynamics we derive here, we show that anomalous terms in both the velocity and the weight rate are nonvanishing already in one dimension, a new feature unique to non-Hermitian systems. We demonstrate this effect by using the non-Hermitian Su-Schrieffer-Heeger model and find good agreement with numerical simulations. Finally, we also consider symmetries to form a complete picture indicating when the anomalous terms might vanish due to novel symmetries unique to non-Hermitian Hamiltonians.

Our work is among only a few that provide a link between the different Berry connections of non-Hermitian bands and physical observables related to transport. While wave-packet dynamics for non-Hermitian systems has been addressed before~\cite{Graefe:2011ba}, and the Berry connection in time-dependent systems has been considered~\cite{Ibanez:2014fv,Wang:2018cx}, only the anomalous velocity has been derived~\cite{Xu:2017bl}, but the full set of equations of motion describing the wave-packet dynamics were not obtained in those works, nor were they adapted to Bloch bands in order to account for dynamics in solid state or other periodic media. By pin-pointing the implications of the various connection terms appearing in the literature~\cite{Garrison:1988dt,Dattoli:1990cp,Sun:1993ip,Shen:2018bx} on wave-packet dynamics, we provide a way to probe them experimentally, e.g., using Bloch oscillations~\cite{Longhi:2015ks,Graefe:2016ey,Longstaff:2019hz}. Given the recent wave of experimental realizations of non-Hermitian systems in engineered metamaterial platforms~\cite{Brandenbourger:2019kf,Helbig:2020bh,Ghatak:2020gf,Hofmann:2020jy}, we are confident that our predictions can be confirmed, and will lead to further insight on the role played by  topology and geometry in non-Hermitian systems. 

The formalism presented here outlines a general path to obtain observables in non-Hermitian systems within semiclassics. Therefore, it may be used to evaluate other quantities beyond those we present here. In contrast with the extensive use of the bi-orthogonal formalism in the literature, we demonstrate how to evaluate actual physical quantities which requires an extension of the conventional analysis in Hermitian systems. 

The remainder of this work is organized as follows:
After introducing conventions and our formalism in Sec.~\ref{sec:formalism}, we explain why the adiabatic theorem fails in non-Hermitian systems and discuss the validity of a single-band approximation in Sec.~\ref{sec:anomalous_velocity}.
In Sec.~\ref{sec:wave_packet_dynamics}, we present the main result of this work, the semiclassical equations of motion for non-Hermitian systems in the presence of an electric field.
Symmetries may impose constraints on the equations of motions, as we discuss in Sec.~\ref{sec:symmetries}.
Finally,in Sec.~\ref{sec:numerics}, we compare our analytical results of the anomalous terms with numerical lattice simulations of the dynamics in one-dimensional systems before concluding in Sec.~\ref{sec:conclusions}.
Several technical details are relegated to the Appendixes.

\section{Formalism and definitions: Berry connection for non-Hermitian systems}
\label{sec:formalism}

In this work, we address a generic linear system that does not necessarily conserve energy due to loss and/or gain. Such a system must be described by the general linear evolution equation
\begin{equation} \label{eq:tdnh}
    i \partial_t \left| \phi(t) \right\rangle = H \left| \phi(t) \right\rangle,
\end{equation}
where $H$ is a non-Hermitian linear operator, $H\neq H^\dagger$, which we will refer to as the ``Hamiltonian.''
Despite this use of terminology, we would like to stress that the system does not have to be quantum mechanical at all. For example, Eq.~\eqref{eq:tdnh} may describe classical diffusion in one dimension when $|\phi(t) \rangle$ is identified with the probability density $P(x,t)$ and $H = iD  \partial_{x}^2$ with $D$ the diffusion constant. It can also describe classical waves with gain or loss; although the wave equation is typically second order in time, it may be decomposed into two first-order equations, which can be cast in the form of Eq.~\eqref{eq:tdnh}. 

Having this in mind, in this work we do not impose any restrictions on the Hamiltonian in terms of symmetry, in particular, we do not restrict ourselves to $PT$-symmetric systems~\footnote{$PT$ symmetry denotes the combination of time-reversal symmetry and inversion symmetry. Under certain conditions that we specify in Appendix~\ref{sec:appendix_symmetries}, the eigenvalues of such a non-Hermitian system are real~\cite{Bender:1998bw,Bender:2007kr,Bender:2010ee}.} and therefore allow the Hamiltonian to have a set of complex energy eigenvalues that we denote by $\varepsilon_\mu$. We consider a time-independent non-Hermitian Hamiltonian such that these energies are the eigenvalues of the time-independent Schr\"{o}dinger equation, 
corresponding to a set of right eigenstates $\ket{ \psi_\mu^R}$, 
\begin{equation}
 H \ket{\psi_\mu^R} = \varepsilon_\mu \ket{ \psi_\mu^R} .
\end{equation}
The right eigenstates are generally different from the left eigenstates of the Hamiltonian that obey
\begin{equation}
 H^\dagger \ket{\psi_\mu^L} = \varepsilon_\mu^* \ket{\psi_\mu^L} .
\end{equation}
For each right eigenstate $\ket{ \psi_\mu^R}$ of $H$ with eigenvalue $\varepsilon_\mu$, there is a corresponding eigenstate $\ket{\psi_\mu^L}$ of $H^\dagger$ with energy $\varepsilon_\mu^*$.
On a lattice (either infinite or with periodic boundary conditions), the states $\ket{ \psi_\mu^R }$ represent Bloch wave functions and can be labeled by their real crystal momentum $\mathbf{k}$ and band index $n$~\footnote{Once boundaries are introduced, it has been suggested that allowing the Bloch wave function to have a complex crystal momentum resolves some issues related to the non-Hermitian skin effect~\cite{Yao:2018cj,Yokomizo:2019fq,Longhi:2019dp,Longhi:2020gy}; we will leave this possibility for future studies.}. Both left and right eigenstates contribute to observables, depending on the measured response~\cite{Schomerus:2020is}.

In non-Hermitian systems, at certain points in parameter space, eigenvalues and eigenvectors of $H$ coalesce, such that there are fewer linearly independent eigenvectors than degrees of freedom~\cite{Heiss:2012bx,Bergholtz2019}, and left and right eigenvectors are orthogonal~\cite{Heiss:2012bx}.
Such points are called ``exceptional points'' and the Hamiltonian is said to be defective.
In this work, we keep away from exceptional or degeneracy points.
Nevertheless, as we show, there are subtleties that enter the construction of the wave packet that stem from crossing points of the imaginary parts of the energies.
When discussing the construction of wave packets, we will comment on these degeneracies in the imaginary part of the energies, whereas we defer the discussion of exceptional points to future work.
Under these conditions, the two sets of eigenstates $\ket{\psi_{n\mathbf{k}}^{R,L}}$ fulfill the following orthonormality conditions~\cite{Brody:2014jv}:
\begin{equation}
 \braket{ \psi^L_{n \mathbf{k}}}{\psi^R_{n^\prime  {\mathbf{k}^\prime}}}
  =\delta_{nn^\prime }\delta_{\mathbf{k}{\mathbf{k}^\prime}} , \label{eq:biorthogonal_states}
\end{equation}
while 
\begin{align}
\braket{\psi^R_{n \mathbf{k}}}{\psi^R_{n^\prime  {\mathbf{k}^\prime}}}
&= I_{nn^\prime } (\mathbf{k}) \delta_{\mathbf{k}{\mathbf{k}^\prime}} ,
\label{eq:overlap_right_eigenstates} \\
\braket{\psi^L_{n \mathbf{k}}}{\psi^L_{n^\prime  {\mathbf{k}^\prime}}}
&= \left[I^{-1} (\mathbf{k}) \right]_{nn^\prime } \delta_{\mathbf{k}{\mathbf{k}^\prime}}
\label{eq:overlap_left_eigenstates}
\end{align}
do not vanish for $n\neq n^\prime $ (although states with different $\mathbf{k}$, either left or right, are still orthogonal due to translational invariance).
The symbol $\delta_{\mathbf{k}{\mathbf{k}^\prime} }$ represents either a Kronecker delta or a $\delta$ function, depending on the boundary conditions.
Here $I (\mathbf{k})$ is the Gramian matrix of the linearly independent right eigenstates, which is positive definite and Hermitian. Its inverse $I^{-1} (\mathbf{k})$ is the Gramian matrix of the left eigenstates.

The Bloch wave function $\psi^{\alpha}_{n\mathbf{k}}(\mathbf{r}) = \braket{ \mathbf{r}}{ \psi^{\alpha}_{n\mathbf{k}}}$ can be split up into a plane-wave contribution and a cell-periodic part, $\psi^{\alpha}_{n\mathbf{k}}(\mathbf{r})=e^{i\mathbf{k}\cdot \mathbf{r} }u^{\alpha}_{n\mathbf{k}}(\mathbf{r})$, with $u^{\alpha}_{n\mathbf{k}} (\mathbf{r}) = \braket{\mathbf{r}}{u^{\alpha}_{n\mathbf{k}}}$ and $\alpha\in R,L$.
We recall that for Hermitian Hamiltonians, the single-band Berry connection and curvature are defined as~\cite{Xiao:2010kw}
\begin{align}
\mathcal{A}_n =& i \braket{ u_{n\mathbf{k}}}{ \partial_\mathbf{k} u_{n\mathbf{k}} } ,\\
\Omega_n =& \nabla_\mathbf{k} \times \mathcal{A}_n .
\end{align}
This form suggests a natural generalization to the space of eigenstates of non-Hermitian Hamiltonians. Instead of a single connection term per band, we now have four, given by~\cite{Garrison:1988dt,Dattoli:1990cp,Shen:2018bx}
\begin{equation}
 \mathcal{A}_n^{\alpha \beta} =
 i \dfrac{\braket{ u_{n\mathbf{k}}^{\alpha}}{ \partial_\mathbf{k} u_{n\mathbf{k}}^{\beta}}}{ \braket{ u_{n\mathbf{k}}^{\alpha}}{u_{n\mathbf{k}}^{\beta}}} =
 \begin{cases}
 i \braket{ u_{n\mathbf{k}}^{\alpha}}{ \partial_{\mathbf{k}} u_{n\mathbf{k}}^{\beta}} & \alpha\neq\beta \\
 i \dfrac{\braket{ u_{n\mathbf{k}}^{\alpha}}{ \partial_\mathbf{k} u_{n\mathbf{k}}^{\alpha}}}{ \braket{ u_{n\mathbf{k}}^{\alpha}}{u_{n\mathbf{k}}^{\alpha}}} & \alpha=\beta
\end{cases}
\label{eq:berry_connection}
\end{equation}
where $\alpha,\beta$ stand for $L,R$.
Pulling the $\mathbf{k}$ derivative around, it is easy to see that these connections are related by
\begin{align}
 \Re \mathcal{A}_n^{RL} &= \Re \mathcal{A}_n^{LR} ,&
 \Im \mathcal{A}_n^{RL} &= -\Im \mathcal{A}_n^{LR} , \nonumber \\
 \Im \mathcal{A}_n^{RR} &= \frac{1}{2} \frac{ \partial_{\mathbf{k}} I_{nn}(\mathbf{k})}{I_{nn}(\mathbf{k})} , &
 \Im \mathcal{A}_n^{LL} &= \frac{1}{2} {\frac{ \partial_{\mathbf{k}} \left[I^{-1} (\mathbf{k}) \right]_{nn}}{ \left[I^{-1} (\mathbf{k}) \right]_{nn} }} \label{eq:reality_conditions} .
\end{align}
We stress that although integrating the resulting four curvatures (using the normalization $I_{nn} (\mathbf{k})=1$ for $\mathcal{A}^{RR}$ and $[I^{-1} (\mathbf{k})]_{nn}=1$ for $\mathcal{A}^{LL}$) over a two-dimensional Brillouin zone (BZ) yields the same topological index~\cite{Shen:2018bx}, their structure is locally different.
We therefore seek to assign a physical meaning to these quantities by highlighting their role in semiclassical wave-packet dynamics.

\section{The single-band wave packet: Hermitian vs.\ non-Hermitian}
\label{sec:anomalous_velocity}

Several underlying assumptions made in standard derivations of the equations of motion governing a wave packet's dynamics become questionable or simply fail when the system is non-Hermitian. Usually, a wave packet is constructed from a single band, such that its propagation in real space is determined by the band's group velocity as well as the single-band Abelian Berry curvature (see Sec.~\ref{sec:wave_packet_dynamics}). This requires energy bands to be resolved, and transitions between bands to be negligible. In other words: These derivations rely on the adiabatic theorem. Once coupling between the bands is introduced via time-dependent terms, it is essential to explore the conditions under which it is possible to treat a single band as isolated. 

The generalization of the adiabatic theorem to evolution under a non-Hermitian Hamiltonian implies that when energies are complex, a particle's ability to stay confined to a single band heavily depends on the imaginary part of the complex energies.
The failure of the standard adiabatic theorem when dealing with non-Hermitian Hamiltonians has been pointed out in the past~\cite{Kvitsinsky:1991km,Nenciu:1992cx}. We repeat the logic here since it is of crucial importance to wave-packet evolution and the topological characterization of the system, and is often overlooked in recent literature.

To be concrete, let us consider a Hamiltonian (not necessarily describing a lattice system) that explicitly changes with time due to some perturbation. In this case, the general ansatz to solve the time-dependent Schr\"{o}dinger equation~\eqref{eq:tdnh} can be expanded in terms of the instantaneous right eigenstates of the Hamiltonian,
\begin{equation}
 \ket{ \phi (t)} = \sum_\mu c_\mu (t) e^{-i \vartheta_\mu (t)} \ket{ \psi_\mu^R(t)} ,
 \label{eq:ansatz_c}
\end{equation}
with complex time-dependent energies, $H(t)\ket{\psi_\mu^R (t)} = \varepsilon_\mu (t) \ket{\psi_\mu^R (t)}$.
We include only the real part of the energies to the phase $\vartheta_\mu (t) =\int_0^t dt^\prime \Re \varepsilon_\mu (t^\prime )$ and take $\hbar=1$, as we do for the remainder of the text.
As we derive explicitly in Appendix~\ref{sec:adiabatic_failure} following Refs.~\onlinecite{Garrison:1988dt,Dattoli:1990cp,Nenciu:1992cx}, the time evolution of the coefficients $c_\mu$ is governed by
\begin{align}
\label{eq:coefficients_time}
 \dot{c}_\mu =& c_\mu ( \Im \varepsilon_\mu - \braket{ \psi_\mu^L }{ \partial_t \psi_\mu^R } ) \\
  &	-\sum_{\nu\neq \mu} c_{\nu} \frac{ \bra{ \psi_\mu^L } (\partial_t H) \ket{\psi_\nu^R}}{\varepsilon_\nu-\varepsilon_\mu} e^{i (\vartheta_\mu -\vartheta_\nu)} , \nonumber
\end{align}
in close analogy with the time evolution in Hermitian systems~\cite{Born:1928hd}.
Equation~\eqref{eq:coefficients_time} is equivalent to the Schr\"{o}dinger equation~\eqref{eq:tdnh}, but recast as a set of coupled equations for the coefficients $c_\mu$ that appear in~\eqref{eq:ansatz_c}.

In Hermitian systems, where left and right eigenvectors are identical and all energies are real, the first term in Eq.~\eqref{eq:coefficients_time} is purely imaginary and thus cannot contribute to any growth or decay of $c_\mu$. The second term describes the coupling to other eigenstates $\nu \neq \mu$. When the Hamiltonian varies slowly with respect to the energy difference between the occupied and the other states, this coupling can be neglected since $\exp [i (\vartheta_\mu-\vartheta_\nu)]$ is a bounded oscillating function~\cite{Born:1928hd}.
This is the standard adiabatic theorem, which implies that a single-band approximation is possible for slowly varying Hamiltonians~\cite{Xiao:2010kw}; if the system starts in an eigenstate, it remains in that eigenstate throughout the evolution. For lattice systems, when applied to spectral flow over the parameter space represented by the lattice momentum, this means that wave-packet dynamics can be confined to a single band, and hence single-band quantities such as the Berry curvature are meaningful objects~\cite{Xiao:2010kw}.

In the non-Hermitian case, however, the first term in Eq.~\eqref{eq:coefficients_time} is generally complex. Its real part $\Im \varepsilon_\mu - \Re \braket{ \psi_\mu^L }{ \partial_t \psi_\mu^R }$ results in an exponential growth or decay of the coefficients $c_\mu$, in stark contrast to the Hermitian case.
Although the coupling to other bands described by the second term in Eq.~\eqref{eq:coefficients_time} has the same functional form as in the Hermitian case, it cannot be safely neglected~\cite{Nenciu:1992cx}: any coupling between bands can be exponentially amplified by the first term.
The imaginary parts of the energies therefore severely affect the ability to neglect such couplings.
If for some $\nu$, $\Im \varepsilon_\mu < \Im \varepsilon_\nu$, the faster exponential growth of the $\nu$th eigenstate will eventually dominate the dynamics of $\ket{\phi (t)}$ and we accordingly cannot consider the $\mu$th eigenstate to be separated from the others---the adiabatic theorem breaks down~\cite{Nenciu:1992cx}.
If we however consider the eigenstate $\mu$ with maximal $\Im \varepsilon_\mu > \Im \varepsilon_\nu$ for all $\nu \ne \mu$, the coefficients $c_{\nu \neq \mu}$ that contribute to Eq.~\eqref{eq:ansatz_c} are exponentially suppressed compared with $c_\mu$. In other words, for a non-Hermitian system, only the state that has the largest gain or smallest loss can be considered to be separated. We will refer to this state as the \emph{dominant} state.

The breakdown of the adiabatic theorem severely affects the semiclassical dynamics of wave packets in lattice systems. In an otherwise time-independent system that is perturbed by an electric field, the crystal momentum $\mathbf{k}$ changes as a function of time and thus serves as the adiabatic parameter. From our previous considerations, we conclude that in order to use the single-band Berry connection to describe the system's dynamics, we must confine our wave packet to the band with the largest imaginary part of the energy, which we label by $n$.
Expanded in terms of its right eigenstates, the wave packet reads
\begin{equation}
 \ket{W (t)} = \sum_{n^\prime } \int_\mathbf{k} w_{n^\prime \mathbf{k}} (t) \ket{\psi^R_{n^\prime \mathbf{k}}} .
\end{equation}
Here, $w_{n^\prime \mathbf{k}} (t) = w_\mathbf{k} (t) \delta_{nn^\prime }$ is the time-dependent envelope function distributing the weight of the wave packet across the BZ, which is assumed to be centered around and rapidly decaying away from the average momentum $\mathbf{k}_c$. This implies that in real space the wave packet is much wider than the lattice spacing.
We denote the properly normalized integral over the whole momentum space by $\int_\mathbf{k}$~\footnote{The normalization depends on the boundary conditions: we either use $\int_\mathbf{k} = \frac{1}{L^d} \sum_\mathbf{k}$ for periodic boundaries, where $L^d$ is the number of lattice sites, or $\int_\mathbf{k} = \frac{1}{(2\pi)^d}\int d^d \mathbf{k}$ for an infinite $d$-dimensional system.}.

Two aspects of non-Hermitian lattice systems impede the construction of wave packets and need to be taken into account.
First, since band energies are not constant as a function of momentum, evolution throughout the BZ generally results in reordering of the imaginary parts of the energies; cf.\ Sec.~\ref{sec:numerics} for some examples.
Hence degeneracy points for the \emph{imaginary parts} of the band energy become crucial---they mark the transition regions in the BZ in which wave packets constructed from eigenstate of a single band switch between bands~\cite{Wang:2018cx,Gong:2019kt}.
Second, the band index $n$ is not necessarily defined globally throughout the whole Brillouin zone. In particular, for so-called point gaps~\cite{Kawabata:2019en} (we will explain this term in Sec.~\ref{sec:numerics}), the index changes upon a translation of $\mathbf{k}$ by a reciprocal-lattice vector. For our purposes it is, however, sufficient to distinguish different bands by an index defined locally in the vicinity of the central momentum $\mathbf{k}_c$, which is always possible.

\section{Wave-packet dynamics in the presence of electric fields}
\label{sec:wave_packet_dynamics}

The gauge-invariant Berry curvature is essential to understand wave-packet dynamics in crystalline materials.
In the presence of external fields, the curvature enters the equations of motion via an ``anomalous velocity'' term. 
For systems in which the dynamics is governed by Hermitian Hamiltonians, a wave packet can be constructed as a superposition of Bloch wave functions from a single band, as long as bands are well-resolved in energy. Defining the average coordinate $\bf{r}_c$, the equations of motion in the presence of an electric field $\bf{E}$ and a magnetic field $\bf{B}$ are given by~\cite{Chang1995,Chang1996,Sundaram:1999ht,Xiao:2010kw}
\begin{align}
\dot{\mathbf{r}}_c = & \partial_{\mathbf{k}} \varepsilon_{\mathbf{k}} |_{\mathbf{k}_c} - \dot{\mathbf{k}}_c\times\mathbf{\Omega}_n,
\label{eq:HermitianEOM1} \\
\dot{\mathbf{k}}_c = & -e \mathbf{E} - e \dot{\mathbf{r}}_c \times \mathbf{B},
\label{eq:HermitianEOM2}
\end{align}
where $e$ is the elementary charge. The last term of Eq.~\eqref{eq:HermitianEOM1} is commonly known as the anomalous velocity term.
It has consequences for transport in two- and higher-dimensional system, and results in a deflection of the particle trajectories~\cite{Xiao:2010kw}.

The semiclassical equations of motion describe the evolution of $\mathbf{r}_c$ and $\mathbf{k}_c$, the center of the wave packet in real and reciprocal space, respectively.
The central position and momentum are expectation values of the corresponding operators with respect to the wave packet.
Generally speaking, the expectation value of an operator $\hat{A}$ is
\begin{equation} \label{eq:expectation}
    \langle A (t) \rangle= \frac{ \bra{W (t)} \hat{A} \ket{W (t)}}{N(t)}.
\end{equation}
In this expression we need to include in the denominator the norm or weight of the wave packet,
\begin{equation}
  N(t)= \braket{W (t)}{W(t)},
\end{equation}
since it can vary in time.
We stress that we define the expectation value as in the Hermitian case, namely with respect to the same state, differently from definitions stemming from the biorthogonal formulation~\cite{Brody:2014jv}.
The biorthogonal formulation relies on a set of left and right eigenstates, such that expectation values are defined with respect to a certain Hamiltonian. Furthermore, even if the state is well-localized around some point in either real or quasimomentum space, the biorthogonal expectation value may be far away from that point and may not even be real.
These issues are avoided by sticking to the standard definition, Eq.~\eqref{eq:expectation}.
This definition also implies that expectation values of Hermitian operators $\hat{A}$ are real. For operators such as position or quasimomentum, Eq.~\eqref{eq:expectation} represents the center of mass of the wave packet in the corresponding space.

The standard route for deriving the equations of motion, outlined for example in Ref.~\onlinecite{Xiao:2010kw}, strongly relies on conservation laws that do not persist when non-Hermiticity is introduced. The construction of a classical Lagrangian and the use of the Euler-Lagrange equations is no longer valid due to the nonconservation of energy and probability density. Therefore, in our case, we must use the Schr\"{o}dinger equation directly to bypass this difficulty, as done, for example, in Ref.~\onlinecite{Lapa:2019bh}. 

The standard equations of motion describe the full derivative of $\mathbf{r}_c$ and $\mathbf{k}_c$ with respect to time. For this, we first define the momentum operator $\hat{\mathbf{k}}$, an operator whose eigenvectors are the Bloch waves and eigenvalues correspond to the crystal momentum $\mathbf{k}$. The expectation value for any function $f$ of $\hat{\mathbf{k}}$ calculated with respect to the wave packet $\ket{W (t)}$ is given by
\begin{align}
 \langle f(\hat{\mathbf{k}}) \rangle
 =& \frac{ \bra{W (t)} f(\hat{\mathbf{k}}) \ket{W (t)}}{ \braket{W (t)}{W(t)}} \\
 =& \frac{1}{N (t)} \sum_{n n^{\prime }} \int_\mathbf{k}  w_{n\mathbf{k}}^* (t) w_{n^\prime \mathbf{k}} (t) I_{nn^\prime } (\mathbf{k}) f(\mathbf{k}) \nonumber,
\end{align}
where we used Eq.~\eqref{eq:overlap_right_eigenstates} for the overlap of the right eigenstates.
The time derivative of any such expectation value therefore requires the evaluation of the time derivative of $w_{n\mathbf{k}} (t)$. Using the Schr\"{o}dinger equation we can write 
\begin{equation}
 \dot{w}_{n\mathbf{k}} = -i w_{n\mathbf{k}} \varepsilon_{n,\mathbf{k}} + i e \sum_{n^\prime } \int_{{\mathbf{k}^\prime}} w_{n^\prime {\mathbf{k}^\prime}} \bra{\psi^L_{n \mathbf{k}}} \Phi (\mathbf{r}) \ket{\psi^R_{n^\prime {\mathbf{k}^\prime}}} ,
\end{equation}
where $\Phi(\mathbf{r})$ is the electrostatic potential that defines the electric field (either physical or artificial, depending on the physical context).
For simplicity, we consider a constant electric field, $\Phi(\mathbf{r})= -\mathbf{E} \cdot \mathbf{r}$, such that
\begin{equation}\label{eq:Wdot}
 \dot{w}_{n\mathbf{k}} = - i  w_{n\mathbf{k}} \varepsilon_{n,\mathbf{k}} - i e \mathbf{E} \cdot \sum_{n'} \int_{{\mathbf{k}^\prime}} w_{n^\prime {\mathbf{k}^\prime}} \bra{\psi_{n\mathbf{k}}^L} \mathbf{r} \ket{\psi_{n^\prime {\mathbf{k}^\prime}}^R} .
\end{equation}  
We note that this form also applies locally in the case where the field is not constant, but varies slowly and smoothly on a length scale that is large compared with the spread of the wave packet. 

While expectation values of the position operator within Bloch bands are known to play a key role in the modern theory of polarization and for Hermitian systems are directly linked to the Berry connection over the BZ, a complete theory generalizing it to non-Hermitian systems is still lacking. To evaluate the matrix element appearing in Eq.~\eqref{eq:Wdot} and similar expressions that include matrix elements of the position operator, we consider the ``weighted matrix element" 
\begin{align}\label{eq:r_expectation}
 & \!\!\!\!\!\!\!\! \int_{\mathbf{k},{\mathbf{k}^\prime} } \bra{\psi_{n \mathbf{k}}^\alpha} \mathbf{r} \ket{\psi_{n^\prime  {\mathbf{k}^\prime}}^\beta} f (\mathbf{k},{\mathbf{k}^\prime}) = i \int_{\mathbf{k},{\mathbf{k}^\prime} } f (\mathbf{k},{\mathbf{k}^\prime}) \nonumber \\
 & \times \left[ \bra{u^\alpha_{n\mathbf{k}}} e^{i \mathbf{r} \cdot ({\mathbf{k}^\prime}-\mathbf{k})} \ket{\partial_{\mathbf{k}^\prime} u^\beta_{n^\prime {\mathbf{k}^\prime}}} - \braket{\psi^\alpha_{n\mathbf{k}}}{\partial_{\mathbf{k}^\prime} \psi^\beta_{n^\prime {\mathbf{k}^\prime}}} \right] \nonumber \\
=& i \int_{\mathbf{k}} \left[ f(\mathbf{k}, \mathbf{k}) \braket{ u_{n\mathbf{k}}^\alpha}{\partial_\mathbf{k} u^\beta_{n^\prime \mathbf{k}}}
+  I_{nn^\prime }^{\alpha \beta} (\mathbf{k}) \left. \partial_{{\mathbf{k}^\prime}} f (\mathbf{k},{\mathbf{k}^\prime}) \right|_{{\mathbf{k}^\prime}=\mathbf{k}}\right] ,
\end{align}
with $I^{\alpha\beta}_{nn^\prime } (\mathbf{k}) = \braket{\psi^\alpha_{n\mathbf{k}}}{\psi^\beta_{n^\prime \mathbf{k}}}$.
To derive the above expression, we used $\mathbf{r} \ket{\psi^\alpha_{n\mathbf{k}}} = i e^{i \mathbf{k} \cdot \mathbf{r}} \partial_\mathbf{k} \ket{u^\alpha_{n\mathbf{k}}}  -i \partial_\mathbf{k} \ket{\psi^\alpha_{n\mathbf{k}}}$ and furthermore that $f(\mathbf{k},{\mathbf{k}^\prime})$ is periodic over the BZ in lattice models or vanishes when either momentum goes to infinity in continuum approximations (e.g., the ubiquitous Dirac approximation).
Equations~\eqref{eq:Wdot} and~\eqref{eq:r_expectation} are sufficient to derive the time derivative of any expectation value that is a function of the momentum operator $\hat{\mathbf{k}}$.
Combining them together, we obtain the following relation
\begin{align}\label{eq:EOM1}
 &\frac{1}{N}
  \frac{d}{dt} \int_{\mathbf{k}} |w_{\mathbf{k}} |^2 I (\mathbf{k}) f (\mathbf{k}) = 2 f (\mathbf{k}_c) \Im \varepsilon_{\mathbf{k}_c} \\
 &- e \mathbf{E} \cdot \left[ \frac{\partial_\mathbf{k} I (\mathbf{k}) |_{\mathbf{k}_c}}{I (\mathbf{k}_c)} f(\mathbf{k}_c)+ \partial_\mathbf{k} f (\mathbf{k})|_{\mathbf{k}_c} + i ( \mathcal{A}^{LR}-\mathcal{A}^{RL})\right] \nonumber 
 \end{align}
for wave packets restricted to one band.
More details on this equation are provided in Appendix~\ref{sec:details_eom}.
Equation~\eqref{eq:EOM1} is one of the central results of this work because it immediately yields two equations of motions. The first is the equation of motion for the evolution of the total weight $N(t)$: plugging in $f(\mathbf{k})=1$, we find that 
\begin{align}\label{eq:dmdt}
\frac{\dot{N}}{N} = 2 \Im \varepsilon_{\mathbf{k}_c} - e \mathbf{E} \cdot \left( \frac{ \partial_\mathbf{k} I (\mathbf{k}) |_{\mathbf{k}_c}}{I (\mathbf{k}_c)} + i(\mathcal{A}^{LR}-\mathcal{A}^{RL})\right) .
\end{align}

We pause here to reflect on the form of Eq.~\eqref{eq:dmdt} that describes the evolution of the total weight of the wave packet. As expected, the total weight diminishes or grows due to the imaginary part of the energy. This effect represents the nonconservation of probability due to the non-Hermiticity of $H$. Less obvious is the interpretation of the second term of this equation, which is directly proportional to the difference in connection terms $\mathcal{A}^{LR}$ and $\mathcal{A}^{RL}$ and the derivative of the Gramian element $I_{nn} (\mathbf{k})$.
It is unusual to find the connection appearing inside an equation of motion, and indeed, in Hermitian systems the anomalous velocity term is proportional to the Berry curvature rather than to the connection itself. In Appendix~\ref{sec:gauge_invariance} we show that this term is gauge invariant, as necessary for an observable.
We defer further discussion of the physical meaning of it to a later stage and first discuss the equations of motion for the centers of the wave packet in momentum and real space. 

Using \eqref{eq:EOM1} with $f(\mathbf{k})= \mathbf{k} $ and supplementing it with \eqref{eq:dmdt} we find that the average momentum evolves according to
\begin{equation}\label{eq:kdot}
\dot{\mathbf{k}}_c = -e \mathbf{E} ,
\end{equation}
as expected.
Any wave packet with a nonzero width in momentum space experiences an additional drift in its central momentum~\cite{Gong:2018ko}, as we discuss in Appendix~\ref{sec:finite_packet_width}.

The derivation of $\dot{\mathbf{r}}_c$ is slightly more involved than that of $\dot{N}$ and $\dot{\mathbf{k}}_c$.
We first note that we can simplify the central position $\mathbf{r}_c$ using Eq.~\eqref{eq:r_expectation}, which gives
\begin{align}
 \mathbf{r}_c 
 &= \left. \mathcal{A}^{RR} (\mathbf{k}) - \frac{i}{2} \frac{\partial_\mathbf{k} I (\mathbf{k})}{I(\mathbf{k})} - \partial_\mathbf{k} \varphi_\mathbf{k} \right|_{\mathbf{k}=\mathbf{k}_c} ,
 \label{eq:central_position}
\end{align}
where $\varphi_\mathbf{k}$ is the (real) phase of the weights, $w_\mathbf{k} = |w_\mathbf{k}| e^{i \varphi_\mathbf{k}}$.
Taking the time derivative of the $i$th coordinate of $\mathbf{r}_c$ gives, once more using Eq.~\eqref{eq:Wdot},
\begin{align}
 (\dot{r}_{c })_i = \partial_{k_i} \Re \varepsilon_\mathbf{k} |_{\mathbf{k}_c} \nonumber
  &- e \sum_j E_j \partial_{k_j} \left(\mathcal{A}_i^{RR}-\frac{i}{2} \frac{ \partial_{k_i} I (\mathbf{k})|_{\mathbf{k}_c}}{I (\mathbf{k}_c)} \right)\\
  &+\frac{1}{2} \sum_j e E_j \partial_{k_i}(\mathcal{A}_j^{LR}+\mathcal{A}_j^{RL}) \label{eq:rdot}.
\end{align}
We give the lengthy but straightforward derivation in Appendix~\ref{sec:details_eom}.

Using Eq.~\eqref{eq:reality_conditions}, we can express the weight rate [Eq.~\eqref{eq:dmdt}] more compactly as
\begin{align}
\frac{\dot{N}}{N} = 2 \Im \left[ \varepsilon_{\mathbf{k}_c} - e \mathbf{E} \cdot \left( \mathcal{A}^{RR}-\mathcal{A}^{LR} \right) \right] .
\label{eq:dmdt_simpler}
\end{align}
Similarly, the velocity [Eq.~\eqref{eq:rdot}] assumes the form
\begin{align}
 (\dot{r}_{c })_i
 =& \partial_{k_i} \Re \varepsilon_\mathbf{k} 
  - e \sum_j E_j \left[ \partial_{k_j} \Re  \mathcal{A}_i^{RR} - \partial_{k_i} \Re \mathcal{A}_j^{LR} \right], \nonumber  \\
 =& \partial_{k_i} \Re \left[ \varepsilon_{\mathbf{k}_c} - e \mathbf{E} \cdot \left( \mathcal{A}^{RR}-\mathcal{A}^{LR} \right) \right] \nonumber \\
  &+ e \left[ \mathbf{E} \times \left( \mathbf{\nabla} \times \Re \mathcal{A}^{RR} \right) \right]_i .
  \label{eq:rdot_simpler}
\end{align}
The last equation reduces to a result derived in Ref.~\onlinecite{Xu:2017bl} if one further imposes $I_{nn}(\mathbf{k})=1$.
Interestingly, both real and imaginary parts of the energy $\varepsilon_{\mathbf{k}}$ appear in these equations in combination with $-e \mathbf{E} \cdot \left( \mathcal{A}^{RR}-\mathcal{A}^{LR} \right)$, leading to the interpretation of the latter as a field-induced correction to the energy in non-Hermitian systems.

Equation~\eqref{eq:rdot_simpler} completes the full set of equations of motion, along with~\eqref{eq:kdot} and~\eqref{eq:dmdt_simpler}.
It contains three terms, the first of which is familiar from Hermitian systems:
The velocity of the wave packet is proportional to the derivative of the real part of the band energy.
If the electric field is zero, this is the only term that contributes to the velocity and it is identical to the equation of motion in the Hermitian case.
The two additional terms are both proportional to the electric field $\mathbf{E}$ and derivatives of the different Berry connection terms.
We therefore identify their sum as the anomalous velocity.
It is easy to verify that this term is gauge-invariant, as we do explicitly in Appendix~\ref{sec:gauge_invariance}; in Appendix~\ref{sec:projectors}, we additionally show how to compute it in a gauge-independent manner using projectors, which is especially useful in numerical calculations.
We also note that when the system is Hermitian, $\mathcal{A}^{RR}=\mathcal{A}^{RL}=\mathcal{A}^{LR}$, Eq.~\eqref{eq:rdot_simpler} reduces to the standard equation of motion upon replacing $\dot{\mathbf{k}}_c = - e \mathbf{E}$,
\begin{equation}
 \dot{\mathbf{r}}_{c } =\partial_{\mathbf{k}} \varepsilon_\mathbf{k}|_{\mathbf{k}_c} - \dot{\mathbf{k}}_c \times \left(\nabla\times\mathcal{A}\right),
\end{equation}
which is identical to Eq.~\eqref{eq:HermitianEOM1}, as expected. 
In contrast, when the system is non-Hermitian, we observe that the anomalous velocity term has a more complicated structure.
We expect the last equation to hold even in the presence of a magnetic field, which would contribute to $\dot{\mathbf{k}}_c$ as in Eq.~\eqref{eq:HermitianEOM2}~\cite{Xu:2017bl}.

It is important to note that in particular, the anomalous velocity generally does not vanish even for one-dimensional systems:
The evolution of $r_c$ assumes the form
\begin{align}
 \dot{r}_{c } = \partial_{k} \Re \varepsilon_k & - e E \partial_{k} \Re \left[ \mathcal{A}^{RR} - \mathcal{A}^{LR} \right] , \label{eq:rdot1D}
\end{align}
suggesting that the connection terms introduced by the band topology in non-Hermitian systems modify the velocity of the wave packet \emph{in the direction of propagation}, along the direction of the electric field. Therefore, we refer to it as the anomalous drift velocity.

\section{Symmetry considerations}
\label{sec:symmetries}

Symmetries play an important role for the semiclassical equations of motion.
In the Hermitian case, time-reversal symmetry constraints the Berry curvature~\cite{Xiao:2010kw}
\begin{equation}
 \Omega_n (-\mathbf{k}) = - \Omega_n (\mathbf{k}) .
\end{equation}
Integrals over the whole BZ therefore vanish in the presence of time-reversal symmetry, which implies, for example, that the Chern number must be zero~\cite{Kane:2005gb}.
Inversion symmetry imposes the additional restriction
\begin{equation}
 \Omega_n (-\mathbf{k}) = \Omega_n (\mathbf{k}) ,
\end{equation}
such that $\Omega_n (\mathbf{k})=0$ when the combination of time-reversal and inversion symmetry is present.
Thus, the anomalous velocity vanishes everywhere in the BZ.

In non-Hermitian systems, we can formally define two different forms of time-reversal symmetry~\cite{Bernard2002,Magnea:2008gd,Kawabata:2019en}.
Time-reversal symmetry may either relate the Hamiltonian to its complex conjugate,
\begin{equation}
 T_+ H_{-\mathbf{k}}^* T_+^\dagger = H_\mathbf{k} 
\end{equation}
or to its transpose
\begin{equation}
 C_+ H_{-\mathbf{k}}^T C_+^\dagger = H_\mathbf{k} 
\end{equation}
with unitary matrices $T_+$ and $C_+$. Both symmetry operations have the same effect on Hermitian operators, but are distinct for non-Hermitian operators.
Combining $T_+$ and $C_+$ gives pseudo-Hermiticity~\cite{Mostafazadeh:2002dq,*Mostafazadeh:2002by,*Mostafazadeh:2002ba}
\begin{equation}
 \eta H_{\mathbf{k}}^\dagger \eta^\dagger = H_\mathbf{k},
\end{equation}
with unitary $\eta$.
Similarly, two different forms of particle-hole symmetry are possible in non-Hermitian systems, which can be combined to two distinct symmetries, namely chiral and sublattice symmetry~\cite{Kawabata:2019en}.
We do not investigate the constraints that all possible antiunitary symmetries impose on anomalous weight rate and velocity, but nevertheless give their explicit form in several cases in Appendix~\ref{sec:appendix_symmetries}. We focus our analysis on $T_+ T_+^* = + 1$ and $C_+ C_+^* = +1$, since the corresponding antiunitary symmetries may relate eigenstates to themselves, differently from $T_+ T_+^* = -1$ and $C_+ C_+^* = -1$ that require two distinct eigenstates at time-reversal invariant momenta. Note that, different from Hermitian systems, the energies of $T_+$ time-reversed partners are not the same, but related via complex conjugation.

The first form of time-reversal symmetry $T_+$ is especially important when (unitary) inversion symmetry
\begin{equation}
 P H_{-\mathbf{k}} P^\dagger = H_\mathbf{k}
\end{equation}
is additionally present, with $P^\dagger P = 1$.
In particular, when $PT_+$ relates each eigenstate at $\mathbf{k}$ to its complex conjugate at the same momentum, $P T_+$ symmetry is \emph{unbroken}~\cite{Bender:1998bw,Konotop:2016cy}, which is only possible when $(P T_+) (P T_+)^* = +1$.
As a consequence, all energy eigenvalues are real~\cite{Bender:1998bw}.

The anomalous velocity term is zero in the presence of unbroken $P T_+$ symmetry.
The relation between the eigenstates and their complex-conjugate constraints the Berry connection [Eq.~\eqref{eq:berry_connection}] and the Gramian matrix [Eq.~\eqref{eq:overlap_right_eigenstates}], as we demonstrate in Appendix~\ref{sec:appendix_symmetries}.
In particular, we find that
\begin{align}
 \mathcal{A}_{n}^{RL} (\mathbf{k}) + \mathcal{A}_{n}^{LR} (\mathbf{k}) = 0 , & &
 \mathcal{A}^{RR} (\mathbf{k}) -\frac{i}{2} \frac{\partial_\mathbf{k} I_{nn} (\mathbf{k})}{I_{nn} (\mathbf{k})} = 0 .
\end{align}
The anomalous contribution to the weight rate, however, may be nonzero in these systems.

In presence of the related $PC_+$ symmetry, similar restrictions apply to the Berry connection and Gramian matrix $I$.
However, neither the anomalous weight rate nor the velocity term are zero because of these symmetry restrictions. Only in Hermitian systems, where $P C_+$ and $P T_+$ symmetries coincide, both contributions are zero.

In the presence of unbroken pseudo-Hermiticity with real energy eigenvalues, both anomalous weight rate and velocity are nonzero. Systems with unbroken pseudo-Hermiticity thus constitute examples of non-Hermitian systems without exponential gain or loss, but with semiclassical dynamics that is fundamentally different from Hermitian systems.
We present a detailed computation of the weight rate and velocity terms in the presence of time-reversal symmetry, inversion symmetry, and pseudo-Hermiticity in Appendix~\ref{sec:appendix_symmetries}.

\section{Numerical simulations}
\label{sec:numerics}

We demonstrate our findings by employing numerical simulations of the wave-packet dynamics in a lattice model.
To this end, we use a non-Hermitian variant of the Su-Schrieffer-Heeger (SSH) Hamiltonian~\cite{Zhu:2014bf,Lee:2016ie,Lieu:2018ff,Kunst:2018ku}
\begin{equation}
 \mathcal{H} (k) = 
 \begin{pmatrix}
  0 & t_L + t_R^\prime  e^{-i k} \\
  t_R + t_L^\prime  e^{i k} & 0
 \end{pmatrix},
 \label{eq:ssh_hamiltonian}
\end{equation}
that describes a one-dimensional (1D) model respecting sublattice symmetry, $\sigma_z \mathcal{H} (k) \sigma_z = -\mathcal{H} (k)$, as well as $PC_+$ symmetry with $P C_+ = \sigma_x$ and $\sigma_x \mathcal{H}^T (k) \sigma_x = \mathcal{H} (k)$~\footnote{Note that both inversion and $C_+$ time-reversal symmetry are broken individually. The combined $PC_+$ symmetry is also referred to as conjugated pseudo-Hermiticity~\cite{Lieu:2018ff}.}.
We set the lattice constant $a=1$ and define for convenience the energy scale $\varepsilon_0 = |t_R|+|t_L|$.

\begin{table}
\centering
{\setlength{\tabcolsep}{1.3em}
\begin{tabular}{c||c|c|c|c}
\toprule
Name & $t_L/\varepsilon_0$ & $t_R/\varepsilon_0$ & $t_L^\prime /\varepsilon_0$ & $t_R^\prime /\varepsilon_0$ \\
 \colrule
(I)   & $11/8$ & $5/8$ & $7/16$ & $-3/8$ \\
(II)  & $12/11$ & $10/11$ & $-5/11$ & $6/11$ \\
(III) & $1.3$ & $0.7$ & $1.6$ & $-0.5$ \\
(IV)  & $1.3i$ & $0.7$ & $0.2$ & $0.9$ \\
\botrule
\end{tabular}}
\caption{Parameters of the non-Hermitian SSH model we use in the numerical simulations.}
\label{tab:parameters}
\end{table}

\begin{figure*}
 \includegraphics{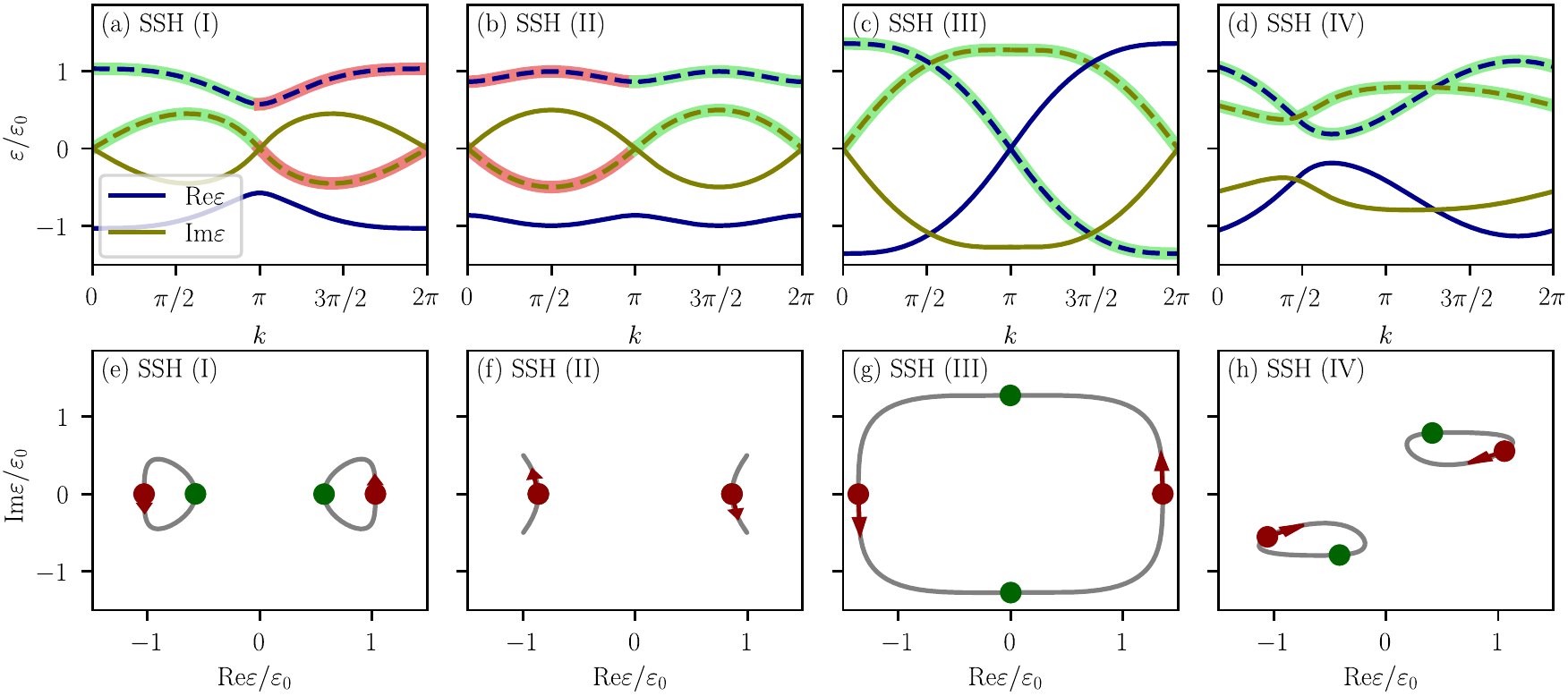}
 \caption{Energy dispersion of the non-Hermitian SSH model~\eqref{eq:ssh_hamiltonian} for the parameter choices specified in Table~\ref{tab:parameters}.
 Panels (a)--(d) show the real (dark blue lines) and imaginary parts (olive lines) of the energies as a function of the crystal momentum $k$. The dashed lines denote the bands we use to construct wave packets whose dynamics we simulate numerically.
 Depending on $k$, the energies of these bands may have the largest imaginary part of both bands (green shaded regions) or the smallest imaginary part (red shaded regions).
 Panels (e)--(h) show the imaginary part of the energies as a function of their real part (gray lines).
 The red and green dots mark the energies at $k=0$ and $k=\pi$, respectively, and the red arrows point from $k=0$ towards positive $k>0$.
 The parameters are chosen such that the models~(I), (II) and~(IV) have a line gap and~(III) has a point gap. Model~(II) is at a fine-tuned point where $\varepsilon_{k}=\varepsilon_{k+\pi}$ coincide, hence each of the two band contours in panel (f) flattens to an arc.}
 \label{fig:energies}
\end{figure*}

The two bands of the SSH model~\eqref{eq:ssh_hamiltonian} are nondegenerate~\cite{Shen:2018bx} apart from isolated points in parameter space.
The model supports both line gaps and point gaps: A line gap implies that the energies in the complex plane do not cross a reference line in the complex-energy plane, whereas a point gap implies that the energies do not cross a reference point~\cite{Kawabata:2019en}.
To demonstrate the universality of our approach, we simulate the wave-packet dynamics using the four parameter choices specified in Table~\ref{tab:parameters}, which cover both line gaps [SSH~(I), (II), and~(IV)] and point gaps [SSH~(III)]; in the latter case the band index switches upon translation by the reciprocal-lattice vector.
None of these cases is close to an exceptional point or a degeneracy point where both real and imaginary energy gaps close at the same momentum.
Both the anomalous weight rate and the anomalous velocity are nonzero and vary with $k$.
In Figs.~\ref{fig:energies}(a)--(d), we show the real and imaginary parts of the energies as a function of momentum, and in Figs.~\ref{fig:energies}(e)--(h), we show the energies in the complex-energy plane.

We simulate the time evolution of a wave packet that is initialized at a certain central position $r_c$ and momentum $k_c$.
We use periodic boundary conditions in all simulations, which is compatible with an electric field implemented as $\mathbf{A} = - \mathbf{E} t$~\footnote{The implementation of the electric field via a position-dependent scalar potential and time-dependent vector potential are related via a gauge transformation, also cf.\ Ref.~\onlinecite{Krieger:1986cn}.}.
Thus, the Hamiltonian $\mathcal{H} (k)\to \mathcal{H} (k+e A(t))$ is time-dependent and we need to discretize time to simulate the evolution of the wave packet via
\begin{equation}
 \ket{ W (t_{n+1}) } = \exp \left[ i (t_{n+1} - t_n) \mathcal{H}_{n+1,n} \right] \ket{ W (t_n) }
\end{equation}
with
\begin{equation}
  \mathcal{H}_{n+1,n} = \mathcal{H} \left(k - e E \frac{t_{n+1}+t_n}{2} \right)
\end{equation}
chosen at times between two time steps.
We prepare the wave packet at $t=0$ by projecting a state $\ket{\tilde{W}_0}$ onto the $n$th band
\begin{equation}
 \ket{W (t=0)} = \sum_{\mathbf{k}} \ket{\psi_{n\mathbf{k}}^R} \braket{\psi_{n\mathbf{k}}^L}{\tilde{W}_0} .
\end{equation}
The state $\ket{\tilde{W}_0}$ is not an eigenstate of the Hamiltonian. It is localized at $k_i=k_c|_{t=0}$ and $r_i = r_c|_{t=0}$ with a momentum-space width $\sigma$ and a real-space width $1/\sigma$. The projected state $\ket{W(t=0)}$ is equally-well localized in real and momentum space since both the projector and $\ket{\tilde{W}_0}$ are diagonal in momentum space, but, different from $\ket{\tilde{W}_0}$, all of the weight of the projected $\ket{W(t=0)}$ is on the $n$th band. Note that we do not require a smooth gauge over the entire Brillouin zone, but only locally around $k_i$, which is always possible, even for topologically nontrivial bands.
In all simulations we employ periodic boundary conditions with period of $L$ lattice sites, and evaluate the central position via $r_c = L/(2\pi) \Im \log \bra{W} \exp(2\pi i \hat{r}/L) \ket{W}$~\cite{Resta:1998hj}.

As we pointed out in Sec.~\ref{sec:anomalous_velocity}, although the standard adiabatic theorem generally does not hold in non-Hermitian systems, it is still possible to stay in the single-band approximation~\cite{Nenciu:1992cx}, which is especially transparent in a two-band model~\cite{Kvitsinsky:1991km,Uzdin:2011dk,Berry:2011ba,Ibanez:2014fv,Milburn:2015hx}.
For sufficiently small electric field and nonzero imaginary part of the energies, the dominant contribution to the weight rate is $\dot{N}/N \approx 2 \Im \varepsilon$ with corrections by the anomalous weight rate that are small compared with $\Im \varepsilon$ for the parameters we consider.
Since sublattice symmetry is present in the two-band SSH model, the energies come in pairs $\pm \varepsilon_{k_c}$, such that whenever $\Im \varepsilon_{k_c} \neq 0$, the weight of one band is exponentially suppressed and the other weight exponentially enhanced.
Although both bands have some nonzero weight for $t>0$, the exponentially enhanced band quickly starts to dominate the wave packet's behavior and the single-band approximation is justified.

\begin{figure}
 \includegraphics{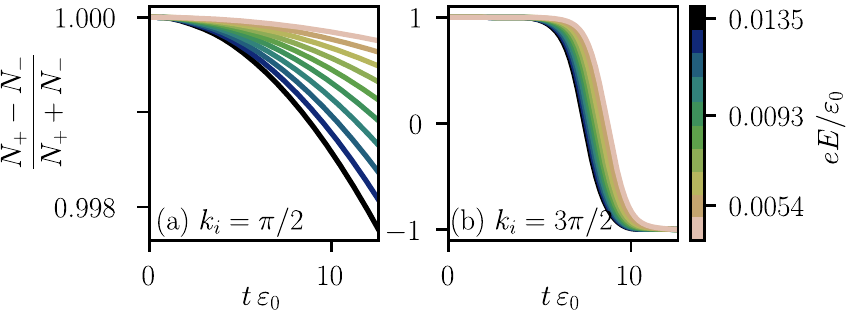}
 \caption{Time evolution of the normalized weight difference $(N_+ - N_-)/(N_+ + N_-)$ for the non-Hermitian SSH model with the parameters chosen as specified in Table~\ref{tab:parameters} [SSH~(I)] and two different initial momenta $k_i = k_c|_{t=0}$:
 (a) $k_i =\pi/2$ with $\Im\varepsilon_{k_i}>0$,
 (b) $k_i =3\pi/2$ with $\Im\varepsilon_{k_i}<0$.
 The different colors denote different strengths of the electric field.
 In all panels, we use periodic boundary conditions with $L=780$ sites, a wave-packet width $\sigma=0.0067$, and discrete time steps with $t_{n+1}-t_n = 0.0168/\varepsilon_0 $.}
 \label{fig:time_evolution}
\end{figure}

We demonstrate this effect in Fig.~\ref{fig:time_evolution}, where we show how the weight is distributed between the two bands of the SSH model.
To compare the weights, we define the single-band weight
\begin{equation}
 N_n = \sum_{\mathbf{k}} I_{nn} (\mathbf{k}) | w_{n\mathbf{k}} |^2,
\end{equation}
where unlike the Hermitian case $N \neq \sum_n N_n$ since mixed terms with $n\neq n^\prime $ also contribute to the total weight of the wave packet.
In Fig.~\ref{fig:time_evolution}(a), we show the evolution of a wave packet's weight that is initialized at $k_c|_{t=0}=\pi/2$ with $\Im \varepsilon_{k_c}|_{t=0} >0$ (in band ``$+$,'' which denotes a positive real part of the energies).
The wave packet remains in its initial band during the whole time evolution (until it eventually reaches momenta with $\Im \varepsilon_{k_c} <0$), with some small leakage to the other band that increases for larger electric fields.
When initializing the wave packet at $k_c|_{t=0}=3\pi/2$ where $\Im \varepsilon_{k_c}|_{t=0} <0$ [Fig.~\ref{fig:time_evolution}(b)], the weight first remains in the initial band.
After some time that is shorter for larger electric fields, the wave packet's weight is distributed over both bands, before the other initially unoccupied band dominates at larger times.

\begin{figure*}
 \includegraphics{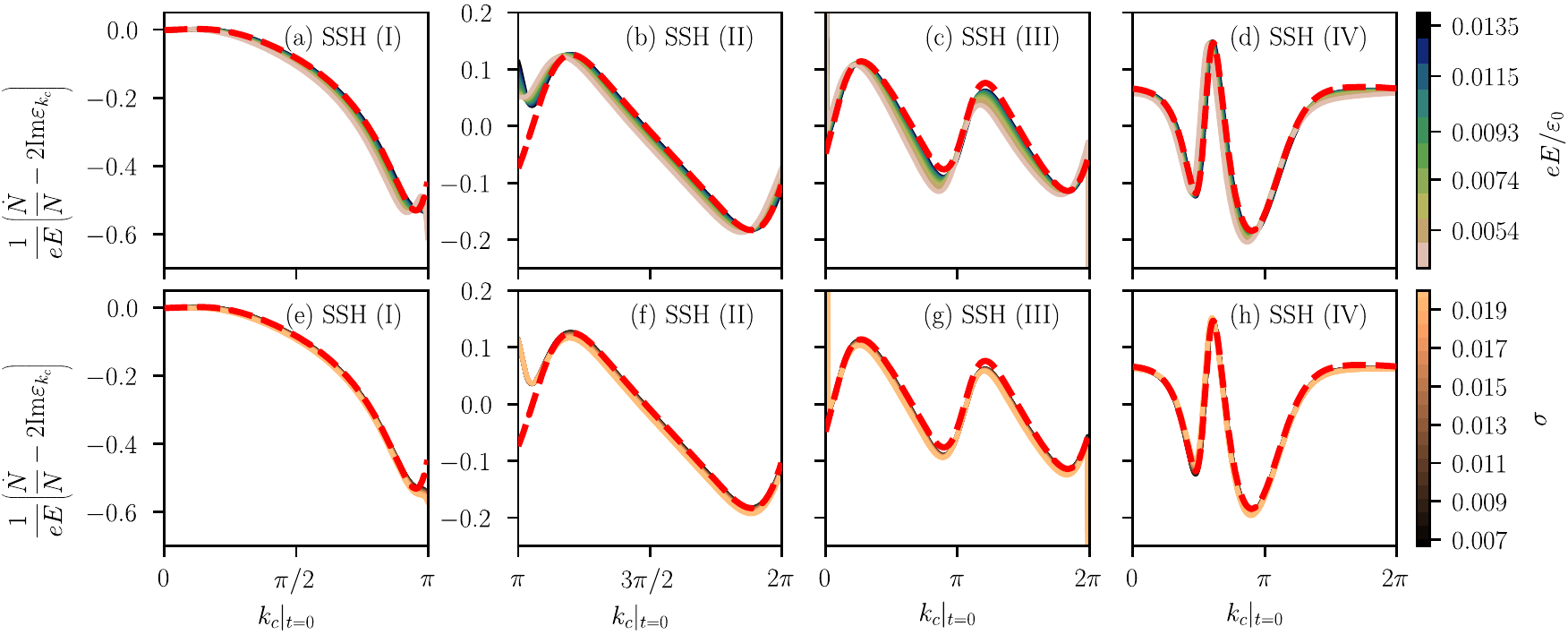}
 \caption{Time evolution of the anomalous weight rate, normalized by the electric field, for the different parameter regimes summarized in Table~\ref{tab:parameters}.
 We extract the anomalous weight rate by subtracting $2\Im \varepsilon_k$ from $\dot{N}/N$ and plot the result divided by the strength of the electric field.
 In panels (a)--(d), we show the resulting anomalous weight rate at fixed wave-packet width ($\sigma=0.0067$) for various strengths of the electric field (with the different panels showing different parameter choices), whereas we compare different wave-packet widths at a fixed electric field ($eE=0.0135\varepsilon_0$) in panels (e)--(h).
 The red dashed lines show the analytic expectation based on Eq.~\eqref{eq:dmdt_simpler} with $eE=0.0135\varepsilon_0$ (the expectation for larger fields is slightly shifted since the central momentum at $t$ depends on the field strength).
 All quantities are evaluated at $t=12.57/\varepsilon_0$. We use $L=780$ sites with periodic boundary conditions and discrete time steps with $t_{n+1}-t_n = 0.0109/\varepsilon_0$ for SSH (III) and $t_{n+1}-t_n = 0.0168/\varepsilon_0$ for the other parameter choices.
 }
 \label{fig:mass}
\end{figure*}

\begin{figure*}
 \includegraphics{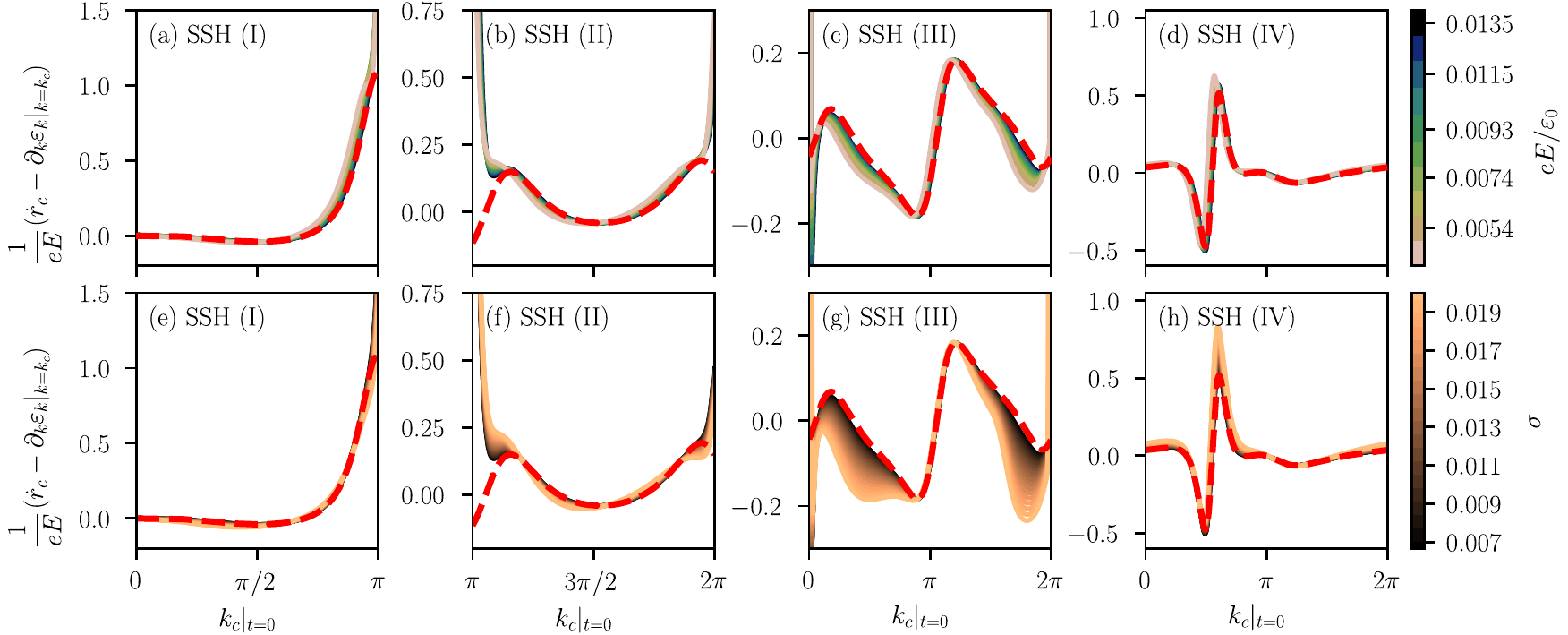}
 \caption{Time evolution of the anomalous velocity for the different parameter regimes summarized in Table~\ref{tab:parameters}.
 We extract the anomalous velocity by subtracting $\partial_k \varepsilon_k |_{k_c}$ from $\dot{r}_c$, and plot the result divided by the strength of the electric field.
 In panels (a)--(d), we show the resulting anomalous velocity at fixed wave-packet width ($\sigma= 0.0067$) for various strengths of the electric field (with the different panels showing different parameter choices), whereas we compare different wave-packet widths at a fixed electric field ($eE=0.0135\varepsilon_0$) in panels (e)--(h).
 The red dashed lines show the analytic expectation based on Eq.~\eqref{eq:rdot_simpler} with $eE=0.0135\varepsilon_0$ (the expectation for larger fields is slightly shifted since the central momentum at $t$ depends on the field strength).
 As in Fig.~\ref{fig:mass}, all quantities are evaluated at $t=12.57/\varepsilon_0$, and we use $L=780$ sites with periodic boundary conditions and discrete time steps with $t_{n+1}-t_n = 0.0109/\varepsilon_0$ for SSH (III) and $t_{n+1}-t_n = 0.0168/\varepsilon_0$ for the other parameter choices.
 }
 \label{fig:velocity}
\end{figure*}

To illustrate our analytical results for the anomalous weight rate and velocity, we compute wave-packet dynamics in the presence of an electric field.
To this end, we simulate the time evolution of the SSH model using various wave-packet widths (ranging from $\sigma=0.0067$ to $\sigma=0.02$), and electric-field strengths (ranging from $eE = 0.00441 \varepsilon_0$ to $eE = 0.0135 \varepsilon_0$).
As we demonstrate in this section, it is necessary to prepare the wave packet such that it is sharply localized in momentum space. Since the localization length $1/\sigma$ in real space must be considerably smaller than the system size to have a well-defined position, the wave packet's momentum-space width is bound from below by the inverse system size. Large system sizes are therefore required for wave packets that are sharply localized in momentum space. We also note after abruptly switching on the electric field that the system relaxes after times $t \sim 1/\varepsilon_0$. The anomalous weight rate and velocity term only start playing a role at times longer than that, thus we focus our analysis on times after this initialization.

We show the numerical results for the anomalous weight rate in Fig.~\ref{fig:mass}. To obtain the anomalous weight rate, we first take the numerical time derivative $\dot{N}/N$ and then subtract the contribution $2\Im \varepsilon_{k_c}$.
The resulting quantity increases approximately linearly with the applied electric field; dividing it by the field gives the Berry-connection induced contribution, Eq.~\eqref{eq:dmdt_simpler}.
In Figs.~\ref{fig:mass}(a)--(d), we show the numerically obtained anomalous weight rate as a function of the initial central momentum $k_c|_{t=0}$ for different electric-field strengths at fixed wave-packet width. The different panels correspond to the different parameter regimes specified in Table~\ref{tab:parameters}.
We only show the evolution for wave packets at initial momenta $k_i = k_c|_{t=0}$ where $\Im \varepsilon_{k_i} \ge 0$, i.e., where the dominant contribution to the wave packet stems from the initial band (cf.\ Fig.~\ref{fig:energies}, where these regions are shaded green).
We compare the anomalous weight rate with the analytical expectation (red dashed lines in Fig.~\ref{fig:mass}), and find generally good agreement.
The only disagreement between the numerical data and the analytical predictions occurs at small momenta [$k\sim \pi$ for SSH~(II) and $k\sim 0$ for SSH~(I) and~(III)]. This is due to the fact that, in this region, time evolution of the central momentum, $\dot{k}_c = - eE$, shifts the wave packet towards smaller $k_c$ where $\Im \varepsilon_{k_c}<0$. There, after some time during which both bands mix, the other band, which is initially unoccupied, starts to dominate (cf.\ Fig.~\ref{fig:time_evolution} for a simulation of the same behavior).
We also note that the quality of agreement between analytical expectation and simulations is sensitive to the model and parameter choices; for example, simulating the model SSH~(III) requires smaller time steps since the energy varies faster as a function of momentum than for the other parameter choices; cf.\ Fig.~\ref{fig:energies}~(c) and~(g).

In Figs.~\ref{fig:mass}(e)--(h), we compare the anomalous weight rate for different wave-packet widths at fixed electric field. The minimal width is chosen such that it is larger than a scale set by the inverse system size $1/L$; for smaller widths, the wave packet is not localized in real space.
The different panels again show the different parameter choices specified in Table~\ref{tab:parameters}.
The agreement between the numerical results and analytical predictions (red dashed lines) is generally better for smaller wave-packet widths, as expected because our analysis relies on wave packets that are sharply localized in momentum space.
As discussed in Appendix~\ref{sec:finite_packet_width}, corrections to $\dot{N}/N$ due to the finite packet width scale with $\sigma^2$.

We show the numerically obtained anomalous velocity in Fig.~\ref{fig:velocity}.
Similarly to the procedure for the anomalous weight rate, we take the numerical time derivative of the central position $\dot{r}_c$, subtract the contribution $\partial_k \varepsilon_k|_{k_c}$ and divide the result by the strength of the electric field to obtain the field-independent Berry-connection induced contribution, Eq.~\eqref{eq:rdot_simpler}.
In Figs.~\ref{fig:velocity}(a)--(d), we compare the anomalous velocity for different strengths of the electric field at fixed wave-packet width.
We again only show those initial momenta where $\Im \varepsilon_{k_i} \ge 0$; cf.\ the green shaded regions in Figs.~\ref{fig:energies}(a)--(d).
The agreement between the numerically obtained velocity and analytical expectation is generally better for larger electric fields, which is an artefact of the numerical simulations: smaller fields require smaller time-step sizes, which are more prone to numerical errors. As we conclude from the data shown Figs.~\ref{fig:velocity}(e)--(h), the agreement is better for smaller wave-packet widths; corrections due to the finite width scale as $\sigma^2$.
Deviations from the expectation are especially pronounced for SSH~(III), but vanish in the limit of infinitesimal wave-packet width.

\section{Summary and conclusions}
\label{sec:conclusions}

In this work, we derived the full set of semiclassical equations of motion that describes wave-packet dynamics in the presence of electric fields for a system governed by a non-Hermitian Hamiltonian and found corrections due to the Berry connections. Instead of two such equations for the center of mass and central momentum, we find three, with the additional equation of motion characterizing the time evolution of the wave packet's weight.
Anomalous terms contribute to the evolution of both the center of mass, namely, the velocity and the weight rate. These terms are proportional to the applied electric field and originate from the Berry connection, i.e., the winding of the eigenstates in momentum space.
Anomalous weight rate and velocity are found to be present already in one-dimensional systems, in marked difference from the Hermitian case.
Our results apply to the band whose energy has the largest imaginary part, for which transitions to other bands are exponentially suppressed~\cite{Nenciu:1992cx}.
As we have shown, transition points between bands follow crossings of the imaginary parts of the complex energies. In the vicinity of such points, the equations of motion cease to be valid, while after transitioning, the single-band description is valid again. Therefore, apart from such degeneracy points (or lines and surface in dimensions greater than one), our equations of motion represent a good description of the dynamics, where the appropriate band into which the wave packet transitions is determined by the reordering of the imaginary part of its energy.

Similar to Hermitian systems, symmetries may constrain the anomalous semiclassical contribution.
We showed that the anomalous velocity vanishes in the presence of unbroken $PT_+$ symmetry, which also implies real energy eigenvalues~\cite{Bender:1998bw,Bender:2010ee}.
This is different from systems with unbroken pseudo-Hermiticity~\cite{Mostafazadeh:2002dq,*Mostafazadeh:2002by,*Mostafazadeh:2002ba} with likewise real energy eigenvalues, for which we showed that both anomalous weight rate and velocity are generally nonzero.
$P C_+$ symmetry, the combination of inversion and another formally possible definition of time-reversal symmetry $C_+$, which coincides with $T_+$ for Hermitian systems~\cite{Kawabata:2019en}, implies neither zero anomalous weight rate nor zero anomalous velocity.

To support our analytical results, we numerically simulated the time evolution of a  wave packet governed by the non-Hermitian SSH model.
These simulations clearly showed the breakdown of the adiabatic theorem for wave packets initialized in bands whose energies do not have the largest imaginary part.
We further isolated the Berry-connection induced anomalous weight rate and velocity and found excellent agreement with the analytical expectation for vanishing wave-packet width.
As we explicitly show in Appendix~\ref{sec:projectors}, both the anomalous weight rate and the velocity (the latter only in one dimension) can be expressed in terms of projectors, which are useful for numerical evaluations since they do not rely on a smooth gauge, different from the expression in terms of Berry connections and Gramian matrix.

We reiterate the importance of our results for one-dimensional systems:
In stark contrast with the Hermitian case, the anomalous velocity is nonzero in one-dimensional systems, which we demonstrated explicitly using the SSH model as an example.
Since its contribution is along the direction of an applied electric field, the anomalous drift  velocity potentially alters the $\sigma_{xx}$ conductivity.
This contribution is naturally inherited from the dissipative nature of the underlying non-Hermitian Hamiltonian.
This is consistent with Ref.~\onlinecite{Xu:2017bl}, where the authors interpreted this term as a modification of the energies~\cite{Xu:2017bl}.
A quantization of the anomalous velocity integrated over the whole Brillouin zone can thus potentially generate a quantized (disorder-independent) contribution to $\sigma_{xx}$ analog to the quantum Hall effect, provided no band switching occurs throughout the BZ. The implications of the diagonal terms on transport are left for future work. 
It would also be interesting to examine the relation between our results and the question of quantized response of open quantum systems described by a Lindblad equation, with topology induced by their coupling to the environment~\cite{Goldstein:2019td,Shavit:2020td,Tonielli:2020em,Yoshida:2019gc}. Another direction to explore is how our results apply to the dynamics of anomalous boundary modes that are described by the long-time limit of a non-Hermitian Hamiltonian~\cite{Lee:2019ft}.

An additional natural extension of our work is the inclusion of magnetic fields into the semiclassical framework. Within the semiclassical approximation, magnetic fields result in a drift in momentum space due to the Lorentz force, i.e., another contribution to $\dot{\mathbf{k}}$~\cite{Xiao:2010kw}.
The effect on non-Hermitian systems may be similar, but we leave determining it to future work, as well as the connection with novel chiral magnetic effects in non-Hermitian systems~\cite{Chernodub:2020dn}.

Our analysis strongly relies on the single-band approximation. This approximation breaks down at degeneracies where bands strongly mix.
In this case, the equations of motion need to be generalized for a subset of bands that remains isolated from the rest of the system.
In Hermitian systems, such a generalization is the non-Abelian Berry connection~\cite{Culcer:2005ke,Shindou:2005ga} that takes into account transitions within the isolated subset of bands.
Moreover, in our analysis we avoided exceptional points, which are unique to non-Hermitian systems~\cite{Heiss:2012bx,Bergholtz2019,Yoshida:2019ca}.
Their treatment is challenging since eigenvalues and eigenvectors coalesce at these points, such that many properties we used break down, e.g., that the matrix of right eigenvectors is invertible.

Going beyond semiclassical dynamics, the anomalous weight rate and velocity terms may be useful for the topological classification of non-Hermitian systems:
Similarly to Hermitian systems, where integrals of the Berry curvature over the whole BZ are quantized in two-dimensional systems, integrals of the Berry-connection induced contribution to the anomalous weight rate and velocity might be quantized in the presence of certain symmetry restrictions.
The relation between this surmised quantization and previously defined topological invariants for non-Hermitian systems~\cite{Lieu:2018ff,Shen:2018bx,Gong:2018ko,Kawabata:2019en,Kawabata:2019gi} is surely of interest.
The anomalous weight rate and velocity may endow a physical interpretation to rather abstract topological invariants, similar to the biorthogonal polarization~\cite{Kunst:2018ku}.

To understand the relation between the anomalous contribution and topology, the role of antiunitary symmetries must be analyzed.
Here we focused on pseudo-Hermiticity and the combination of inversion symmetry with the two formally possible definitions of time-reversal symmetry, where eigenvectors are related by a reality condition.
Particle-hole symmetries may impose similar restrictions on the eigenstates because they can, as opposed to Hermitian systems, relate eigenstates to themselves.
A careful analysis of the interplay of symmetries and anomalous contributions to the semiclassical equations of motion can give rise to a deeper understanding of the role of symmetries and topology in non-Hermitian systems.

Finally, incorporating boundaries into the semiclassical equations of motion is a challenging task. In non-Hermitian systems, it has been suggested that allowing for complex crystal momenta potentially incorporates boundary states of open systems~\cite{Yao:2018cj,Yokomizo:2019fq,Longhi:2019dp,Longhi:2020gy}. It would be interesting to generalize our approach to complex momenta, which is not only relevant for open systems, but also has the potential to give hints how to resolve questions related to the bulk-boundary correspondence in non-Hermitian systems.

\begin{acknowledgments}
The authors are grateful to N.~Regnault for his insightful comments and careful reading of the paper, and to Y.~Xu for bringing their previous work~\cite{Xu:2017bl} to our attention, where the anomalous velocity has been derived. We also thank L.~Beilkin-Sirota, B.~A.~Bernevig, J.~C.~Budich, A.~Cortijo, A.~G.~Grushin, S.~Huber, Y.~Lahini, V.~Peri, and Y.~Shokef for useful discussions.
J.~B.\ is supported by the ERC Starting Grant No.~678795 TopInSy.
M.~G.\ and R.~I.\ are supported by the Israel Science Foundation (ISF, Grants No.~227/15 and 1790/18, respectively) and by the US-Israel Binational Science Foundation (BSF, Grants No.~2016224 and 2018226, respectively).
\end{acknowledgments}

\appendix

\section{Failure of the adiabatic theorem for non-Hermitian Hamiltonians}
\label{sec:adiabatic_failure}

For completeness, we repeat the standard derivation of the adiabatic theorem~\cite{Born:1928hd}, but take into account that our Hamiltonian is non-Hermitian.
We investigate how states governed by a time-dependent non-Hermitian Hamiltonian evolve in time and show why and when the adiabatic theorem fails.
Since the Hamiltonian under consideration is non-Hermitian, its left and right instantaneous eigenstates are generally different and satisfy
\begin{align}
 H (t) \ket{\psi_\mu^R (t)} = \varepsilon_\mu (t) \ket{ \psi_\mu^R (t) }, & &
 H^\dagger (t) \ket{\psi_\mu^L} = \varepsilon_\mu^* (t) \ket{\psi_\mu^L (t) }
\end{align}
with complex time-dependent energies $\varepsilon_\mu (t)$.
To solve the time-dependent Schr\"{o}dinger equation~\eqref{eq:tdnh}, we use an ansatz $\ket{\phi (t)}$ that we express in terms of the right instantaneous eigenstates of $H(t)$,
\begin{equation}
 \ket{ \phi (t)} = \sum_\mu c_\mu (t) e^{-i \vartheta_\mu (t)} \ket{ \psi_\mu^R (t)} ,
\end{equation}
where we defined the phase $\vartheta_\mu (t)=\int_0^t dt^\prime \Re \varepsilon_\mu (t^\prime )$ for convenience. 
From $[i \partial_t -H(t) ] \ket{\phi (t)} = 0$ we get (dropping the label $t$ and keeping in mind that all quantities depend on time)
\begin{align}
 \sum_\mu e^{-i \vartheta_\mu} \left[ i \dot{c}_\mu  + c_\mu \partial_t \vartheta_\mu + i c_\mu  \partial_t - c_\mu H \right] \ket{\psi_\mu^R} =0 .
\end{align}
Using $H \ket{ \psi_\mu^R }= \varepsilon_\mu \ket{ \psi_\mu^R}$ and $\partial_t\vartheta_\mu= \Re \varepsilon_\mu$ gives
\begin{equation}
 \sum_\mu e^{-i \vartheta_\mu } \left[ \dot{c}_\mu - c_\mu \Im \varepsilon_\mu + c_\mu \partial_t \right] \ket{ \psi_\mu^R} = 0 ,
\end{equation}
which can be rewritten as
\begin{equation}
\sum_\mu \dot{c}_\mu e^{-i \vartheta_\mu} \ket{ \psi_\mu^R } = \sum_\mu c_\mu e^{-i \vartheta_\mu}(\Im \varepsilon_\mu -\partial_t) \ket{ \psi_\mu^R } .
\label{eq:appendix_cevolution}
\end{equation}
At this point, we utilize the orthogonality between left and right eigenvectors away from exceptional points, namely $\braket{ \psi_\mu^L}{ \psi_\nu^R} =\delta_{\mu\nu}$, assuming a proper normalization of left and right eigenstates. Multiplying Eq.~\eqref{eq:appendix_cevolution} from the left with $\bra{\psi_\nu^L}$ gives
\begin{equation}\label{eq:coefficients}
 \dot{c}_\nu = c_\nu \Im \varepsilon_\nu -\sum_\mu c_\mu \braket{ \psi_\nu^L}{ \partial_t \psi_\mu^R} e^{i (\vartheta_\nu(t)-\vartheta_\mu(t))} .
\end{equation}
We bring the above expression for the coefficients into a different form by first taking the time derivative of the Schr\"{o}dinger equation
\begin{equation}
 \partial_t ( H \ket{\psi_\mu^R} ) =  (\partial_t H) \ket{\psi_\mu^R} + H \ket{ \partial_t \psi_\mu^R}
\end{equation}
and then multiplying from the left by $\bra{\psi_\nu^L}$ with $\nu \neq \mu$, again using that left and right eigenstates are orthonormal, 
\begin{equation}
 (\varepsilon_\mu -\varepsilon_\nu) \braket{\psi_\nu^L}{ \partial_t \psi_\mu^R}
 = \bra{\psi_\nu^L} (\partial_t H) \ket{\psi_\mu^R} .
\end{equation}
Separating the two cases $\mu\neq \nu$ and $\mu=\nu$ in Eq.~\eqref{eq:coefficients} gives
\begin{align}
\label{eq:coefficent_ad}
  \dot{c}_\mu =& c_\mu (\Im \varepsilon_\mu - \braket{ \psi_\mu^L}{\partial_t \psi_\mu^R}) \\
&-\sum_{\nu \neq \mu} c_\nu \frac{ \bra{ \psi_\mu^L} ( \partial_tH) \ket{ \psi_\nu^R} }{\varepsilon_\nu-\varepsilon_\mu} e^{i (\vartheta_\mu -\vartheta_\nu)} \nonumber,
\end{align}
where we exchanged $\mu \leftrightarrow \nu$ and assumed that the energies are nondegenerate.

We pause here to note that if we had considered a Hermitian Hamiltonian where the left and right sets coincided and the energies were real, we would want to claim that the second term on the right-hand side was negligible for slow variations of the Hamiltonian (with respect to the energy difference between the occupied level and the rest).
This is the standard adiabatic theorem~\cite{Born:1928hd} and it allows dropping the second term, provided the bands are separated such that the denominator is finite, and the exponent that tags along is a bounded function (which is indeed the case, since it is oscillatory).

In our case, however, the energies are not real, and therefore there is a more stringent condition on when the second term can be neglected.
To understand the implications of Eq.~\eqref{eq:coefficent_ad} on non-Hermitian systems, we note that the first term $\Im \varepsilon_\mu - \braket{ \psi_\mu^L}{\partial_t \psi_\mu^R}$ is generally complex, different from Hermitian systems where it is purely imaginary (since $\Im \varepsilon_\mu =0$ and the left and right eigenvectors coincide).
This results in a phase rotation due to its imaginary part and additionally to an exponential growth or decay of the amplitude $|c_\mu|$ due to its real part.

The following discussion applies to systems governed by a slowly varying Hamiltonian whose energy eigenvalues have imaginary parts with absolute values much larger than the Berry connection, $|\Im \varepsilon_\mu | \gg |\Re \braket{ \psi_\mu^L}{\partial_t \psi_\mu^R}|$, such that the energy eigenvalues dominate the dynamics of $\ket{\phi (t)}$.
Consider the time evolution of a mixed state where all the weight is initially equally distributed over all $c_\mu$.
The state with the largest imaginary part of the energies grows the fastest/decays the slowest and thus dominates the state's behavior when coupling between the eigenstates can be neglected.
In this particular situation, such coupling is negligible when the Hamiltonian varies slowly compared with the energy difference between the different states.

Now consider a state prepared in a single eigenstate $\mu$ with coefficients $c_\mu (t=0) = 1$ and $c_{\nu\neq \mu} (t=0) =0$. The real and imaginary parts of $\bra{ \psi_\mu^L} ( \partial_tH) \ket{ \psi_\nu^R} /(\varepsilon_\nu-\varepsilon_\mu)$ define inverse time scales that become smaller the more slowly the Hamiltonian varies. Due to the oscillating term $\exp[ i (\vartheta_\mu -\vartheta_\nu)]$, integration over time yields a bounded result. Although this coupling term is bound, the coefficient $c_{\nu \neq \mu}$ can still be relevant due to their exponential growth originating in $\dot{c}_\nu \sim c_\nu \Im \varepsilon_\nu$---this is strikingly different from the Hermitian case where the only growth of the amplitude of $c_\nu$ is due to coupling to other bands.
Thus, even for Hamiltonians that vary slowly compared with the gap, coupling to unoccupied eigenstates may play an important role in the system's dynamics~\cite{Kvitsinsky:1991km,Nenciu:1992cx}.

When initializing a state in the eigenstate $\mu$ with the largest imaginary part $\Im \varepsilon_\mu > \Im \varepsilon_\nu$ for all $\nu \neq \mu$, which corresponds to the largest gain or smallest loss, the adiabatic theorem still holds for the $\mu$th eigenstate. Any exponential growth of another eigenstate $\nu \neq \mu$ is always overwhelmed by $\Im \varepsilon_\mu > \Im \varepsilon_\nu$. Thus, the adiabatic theorem holds; this has been proven more rigorously in Ref.~\onlinecite{Nenciu:1992cx}.

In the case of a lattice system that we consider throughout most of the main text, the energies are band energies, and thus will generally depend on the crystal momentum $\mathbf{k}$ (changing the label $\mu \to n,\mathbf{k}$ to band index and momentum).
Therefore it is possible for a band to be dominant and for the adiabatic theorem to hold only for some regions of the BZ and not for all of it. This has a crucial effect when we consider semiclassics and the equations of motion: Due to the evolution of $\mathbf{k}$ via $\dot{\mathbf{k}} = - e \mathbf{E}$, the imaginary parts reorder in time, such that an eigenstate initialized in a band with the largest imaginary part is suppressed by another band as soon as the imaginary parts of their band energies cross. We provide examples for such crossings and corresponding reordering of band energies in Sec.~\ref{sec:numerics} in the main text.

\section{Corrections due to finite wave-packet width}
\label{sec:finite_packet_width}

\begin{figure}
 \includegraphics{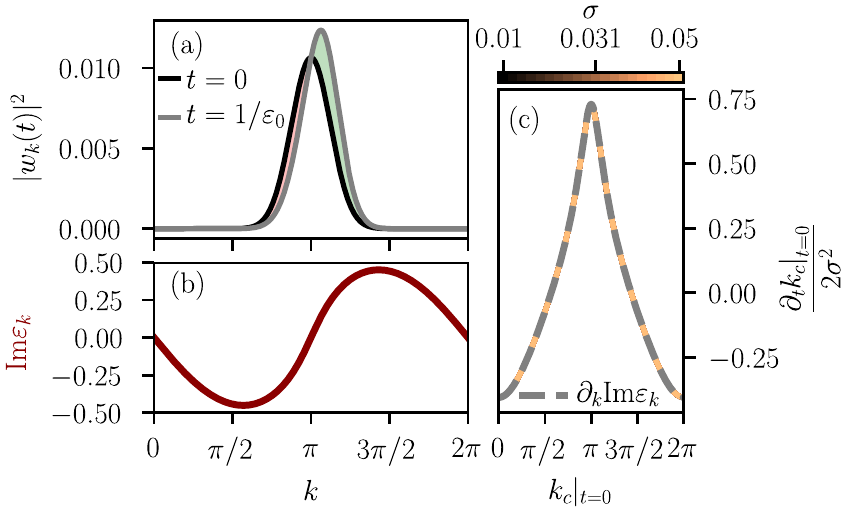}
 \caption{The shift of a wave packet's central momentum due to its nonzero width in momentum space. (a) Without an electric field, the absolute value of the weights $w_k (t)$ increases or decreases in time, depending on the sign of the imaginary part of the energies [cf. panel (b)]. A wave packet initialized at $k_c|_{t=0}=0$ (black line) loses weight for $k<0$ (red shaded area) and gains weight for $k>0$ (green shaded area) when evolving in time, such that its shape and central momentum changes (gray line).
 (c) We show $\partial_t k_c |_{t=0}$ for different widths $\sigma$ that collapse onto the same line when rescaling it by $2\sigma^2$. The rescaled value equals $\partial_k \Im \varepsilon_k$ (gray dashed line); cf.\ Eq.~\eqref{eq:time_derivate_momentum} and note that $\Im \varepsilon_{k_c}=0$.
 In all simulations, we employ the SSH model [Eq.~\eqref{eq:ssh_hamiltonian}] with the parameters specified in Table~\ref{tab:parameters} [SSH~(I)]. We use $L=600$ lattice sites and periodic boundary conditions.}
 \label{fig:gain_loss}
\end{figure}

Any wave packet that is not delocalized in real space has a certain nonzero width $\sigma$ in momentum space.
For non-Hermitian Hamiltonians this width automatically induces a time-dependent shift of the wave packet's central momentum, as we demonstrate in this appendix (see also Ref.~\onlinecite{Gong:2018ko}).
For simplicity we will explicitly treat the 1D case, although the results for higher dimensions are similar.
One easy way to understand this shift in momentum space is to consider the time evolution of a wave packet centered at $k_c=0$ governed by a Hamiltonian with $\Im \varepsilon_k |_{k=0} = 0$, but $\partial_k \Im \varepsilon_k |_{k=0} >0$, such that the imaginary part of the energies is (in the vicinity of $k=0$) positive for $k>0$ and negative for $k<0$.
When evolving the wave packet in time, it gains weight for $k>0$ and loses weight for $k<0$. This gives an effective shift of the central momentum $k_c$ in time.
In Fig.~\ref{fig:gain_loss} we demonstrate this behavior by using the numerically obtained time evolution of the non-Hermitian SSH model, Eq.~\eqref{eq:ssh_hamiltonian}.
In panel~(a) we show the squared weights $|w_{k}|^2$ of a wave packet [black line in panel (a)] initialized in a single band with negative imaginary part of the energies for $k<0$ and positive $\Im \varepsilon_k>0$ for $k>0$; cf.\ panel~(b).
To demonstrate the gain and loss of the weights, we compare the initial wave packet with the time-evolved wave packet at time $t=1/\varepsilon_0$ [gray line in panel~(b)]; the wave packet loses weight for $k<0$ and gains weight for $k>0$, as expected.
As a result, the central momentum shifts with time to $k_c>0$, and hence $\partial_t k_c >0$.

A more formal way to understand the shift in momentum is by a Taylor expansion around the central momentum $k_c$.
Consider the integral
\begin{align}
 \int d k \,& |w_k|^2 I (k) f (k) \approx \int d k \,|w_k|^2 I (k) \left[ f (k_c)  \right. \\
 & \left. + \partial_k f (k) |_{k=k_c} ( k-k_c)+  \frac{1}{2} \partial_k^2 f (k) |_{k=k_c} ( k-k_c)^2 \right] \nonumber ,
\end{align}
where we expanded $f(k)$ around $k_c$.
By employing the definition of the central momentum $\int d k \,|w_k |^2 I (k) ( k-k_c) = 0$ and the second moment $\int d k \,|w_k|^2 I_{nn} (k) ( k-k_c)^2 = N \sigma^2$, we obtain, using that the weight is sharply peaked around $k_c$~\cite{Gong:2018ko},
\begin{align}
  \int d k \,& |w_k|^2 I (k) f (k) \approx N f (k_c) + \frac{1}{2}  N \sigma^2 \partial_k^2 f (k) |_{k=k_c} .
 \label{eq:finite_width}
\end{align}

In the absence of an electric field, we employ Eq.~\eqref{eq:finite_width} to determine corrections to the norm rate and velocity.
In particular, we have
\begin{equation}
  \frac{\dot{N}}{N} = \frac{2}{N} \int dk \,|w_k|^2 I (k)  \Im \varepsilon_k
  \approx  2 \Im \varepsilon_{k_c} + \sigma^2 \partial_k^2 \Im \varepsilon_k |_{k_c} .
\end{equation}
We further evaluate the time derivative of unnormalized momentum expectation value,
\begin{align}
 \frac{d}{dt}
 & \bra{W} \hat{k} \ket{W} = 2 \int dk \,|w_k|^2 I (k) k \Im \varepsilon_k \\
 & \approx N \left[ 2 k \Im \varepsilon_{k} + \sigma^2 \left( k \partial_k^2 \Im \varepsilon_k + 2 \partial_k \Im \varepsilon_k \right) \right]_{k=k_c} \nonumber ,
\end{align}
such that the time derivative of the central momentum
\begin{align}
 \dot{k}_c
 =& \frac{1}{N} \left[ \frac{d}{dt} \bra{W} \hat{k} \ket{W} - N k_c \frac{\dot{N}}{N} \right] \\
 =& 2 \sigma^2 \left. \partial_k \Im \varepsilon_k \right|_{k=k_c} ,
 \label{eq:time_derivate_momentum}
\end{align}
This expectation agrees well with the numerical simulations, as we demonstrate in Fig.~\ref{fig:gain_loss}(c). There we compare the numerically obtained time derivative $\partial_t k_c |_{t=0}$ for the non-Hermitian SSH model [Eq.~\eqref{eq:ssh_hamiltonian}] with the momentum derivative of the imaginary part of the energies $\partial_k \Im \varepsilon_k |_{k=k_c(t=0)}$ and find that the data matches well the prediction, Eq.~\eqref{eq:time_derivate_momentum}.

\section{Details on the derivation of the equations of motion}
\label{sec:details_eom}

In the main text, we sketched how central relations between the expectation value of the position operator and the different Berry connections give rise to the semiclassical equations of motion that govern the wave-packet dynamics.
Here, we provide more details on these relations, in particular, which assumptions were used in their derivation.
We start by deriving Eqs.~\eqref{eq:r_expectation} and~\eqref{eq:EOM1} before turning to the position operator.

\subsection{Momentum-dependent functions}

The position operator acting on a Bloch eigenstate $\mathbf{r} \ket{\psi^\alpha_{n\mathbf{k}}}$ can be expressed by derivatives of the Bloch and cell-periodic eigenstates,
\begin{equation}
\mathbf{r} \ket{\psi^\alpha_{n\mathbf{k}}} = i e^{i \mathbf{k} \cdot \mathbf{r}} \partial_\mathbf{k} \ket{u^\alpha_{n\mathbf{k}}}  -i \partial_\mathbf{k} \ket{\psi^\alpha_{n\mathbf{k}}} ,
\label{eq:position_bloch_eigenstate}
\end{equation}
such that the overlap (with $\alpha,\beta \in L,R$)
\begin{align}
\bra{\psi^\alpha_{n\mathbf{k}}} \mathbf{r} \ket{\psi^\beta_{n^\prime {\mathbf{k}^\prime}}}
 =& i \bra{u^\alpha_{n\mathbf{k}}} e^{i \mathbf{r} \cdot ({\mathbf{k}^\prime}-\mathbf{k})} \ket{\partial_{\mathbf{k}^\prime} u^\beta_{n^\prime {\mathbf{k}^\prime}}}  \nonumber \\
 &-i \braket{\psi^\alpha_{n\mathbf{k}}}{\partial_{\mathbf{k}^\prime} \psi^\beta_{n^\prime {\mathbf{k}^\prime}}} .
\end{align}
Integrating the overlap multiplied by a function $f (\mathbf{k},{\mathbf{k}^\prime})$ over both momenta $\mathbf{k}$ and ${\mathbf{k}^\prime}$ yields
\begin{align}
 \int_{\mathbf{k},{\mathbf{k}^\prime}}  f (\mathbf{k},{\mathbf{k}^\prime}) & \bra{u^\alpha_{n\mathbf{k}}} e^{i \mathbf{r} \cdot ({\mathbf{k}^\prime}-\mathbf{k})} \ket{\partial_{\mathbf{k}^\prime} u^\beta_{n^\prime {\mathbf{k}^\prime}}}
  \nonumber \\
 &= \int_{\mathbf{k}} f (\mathbf{k},\mathbf{k}) \braket{u^\alpha_{n\mathbf{k}}}{\partial_\mathbf{k} u^\beta_{n^\prime \mathbf{k}}},
\end{align}
where we used that the functions $u_{n\mathbf{k}}^\alpha (\mathbf{r})$ are periodic in real space with the lattice period, hence the inner product $\bra{u^\alpha_{n\mathbf{k}}} e^{i \mathbf{r} \cdot ({\mathbf{k}^\prime}-\mathbf{k})} \ket{\partial_{\mathbf{k}^\prime} u^\beta_{n^\prime {\mathbf{k}^\prime}}}$, which is a Fourier transform over real space, vanishes unless $\mathbf{k}-{\mathbf{k}^\prime}$ equals a reciprocal-lattice vector, which needs to be zero since both momenta are in the first BZ.
The second integral is
\begin{align}
 \int_{{\mathbf{k}^\prime}} f (\mathbf{k},{\mathbf{k}^\prime}) \braket{\psi^\alpha_{n\mathbf{k}}}{\partial_{\mathbf{k}^\prime} \psi^\beta_{n^\prime {\mathbf{k}^\prime}}}
  =& -  I^{\alpha\beta}_{nn^\prime } (\mathbf{k}) \partial_{\mathbf{k}^\prime} f(\mathbf{k},{\mathbf{k}^\prime}) |_{{\mathbf{k}^\prime}=\mathbf{k}} ,
\end{align}
with $I^{\alpha\beta}_{nn^\prime } (\mathbf{k}) =\braket{\psi^\alpha_{n\mathbf{k}}}{\psi^\beta_{n^\prime \mathbf{k}}}$.
For the second relation, we assumed that $f(\mathbf{k},{\mathbf{k}^\prime})$ is periodic in $\mathbf{k}$ for lattice systems, or vanishes when one of the momenta goes to infinity in continuous systems.
Combining both relations  gives
\begin{align}
 \int_{\mathbf{k},{\mathbf{k}^\prime}} f(\mathbf{k},{\mathbf{k}^\prime})\bra{\psi^\alpha_{n\mathbf{k}}} \mathbf{r}  &  \ket{\psi^\beta_{n^\prime {\mathbf{k}^\prime}}}
 = i \int_\mathbf{k} \left[ f (\mathbf{k},\mathbf{k}) \braket{u^\alpha_{n\mathbf{k}}}{\partial_\mathbf{k} u^\beta_{n^\prime \mathbf{k}}} \right. \nonumber \\
 & \left. + I_{nn^\prime }^{\alpha\beta} (\mathbf{k}) \left. \partial_{{\mathbf{k}^\prime}} f (\mathbf{k},{\mathbf{k}^\prime}) \right|_{{\mathbf{k}^\prime}= \mathbf{k}} \right] ,
 \label{eq:r_expectation_appendix}
\end{align}
as quoted in the main text, Eq.~\eqref{eq:r_expectation}.

We use Eq.~\eqref{eq:r_expectation_appendix} to derive the time derivative of any operator that is a function of the momentum operator $\hat{\mathbf{k}}$; in particular,
\begin{align}
 \frac{1}{N} \frac{d}{dt} \bra{W(t)} f (\hat{\mathbf{k}}) \ket{W (t)}
 =&  \frac{1}{N}  \sum_{nn^\prime } \int_{\mathbf{k}} f(\mathbf{k}) I_{nn^\prime } (\mathbf{k})  \\
  & \times \left( \dot{w}_{n\mathbf{k}}^* w_{n^\prime \mathbf{k}} + w_{n\mathbf{k}}^* \dot{w}_{n^\prime \mathbf{k}} \right) \nonumber  ,
\end{align}
where the time evolution of $w_{n\mathbf{k}}$ is governed by Eq.~\eqref{eq:Wdot}.
We now assume that the single-band approximation holds, i.e., that the weight $|w_{n\mathbf{k}}|^2$ is concentrated in a single band, such that we can drop the band indices and do not have to consider cross terms with $n\neq n^\prime $.
Inserting the time derivatives $\dot{w}_{\mathbf{k}}^*$ and $\dot{w}_{\mathbf{k}}$ gives
\begin{align}
 & \frac{d}{dt} \bra{W(t)} f (\mathbf{k}) \ket{W (t)}
 = i \int_{\mathbf{k}} f(\mathbf{k}) I (\mathbf{k}) \left[ |w_{\mathbf{k}} |^2  ( \varepsilon_{\mathbf{k}}^* - \varepsilon_{\mathbf{k}} ) \right. \nonumber \\
  & \left. + e \mathbf{E} \cdot \int_{{\mathbf{k}^\prime}} \left( w_{{\mathbf{k}^\prime}}^* w_{\mathbf{k}} \bra{\psi^R_{{\mathbf{k}^\prime}}} \mathbf{r} \ket{\psi^L_{\mathbf{k}}} - w_{\mathbf{k}}^* w_{{\mathbf{k}^\prime}} \bra{\psi_{\mathbf{k}}^L} \mathbf{r} \ket{\psi^R_{{\mathbf{k}^\prime}}} \right) \right] .
\end{align}
We use Eq.~\eqref{eq:r_expectation_appendix} to evaluate the double integral over $\mathbf{k},{\mathbf{k}^\prime}$,
\begin{align}
\int_{\mathbf{k},{\mathbf{k}^\prime}} f I & \left( w_{{\mathbf{k}^\prime}}^* w_{\mathbf{k}} \bra{\psi^R_{{\mathbf{k}^\prime}}} \mathbf{r} \ket{\psi^L_{\mathbf{k}}} - w_{\mathbf{k}}^* w_{{\mathbf{k}^\prime}} \bra{\psi_{\mathbf{k}}^L} \mathbf{r} \ket{\psi^R_{{\mathbf{k}^\prime}}} \right) \nonumber \\
 =& \int_{\mathbf{k}} |w_\mathbf{k}|^2  \left[ f I \left( \mathcal{A}^{RL} - \mathcal{A}^{LR} \right) + i \partial_\mathbf{k} \left( f I \right) \right] ,
\end{align}
such that the time derivative evaluates to
\begin{align}
 \frac{1}{N} \frac{d}{dt} \bra{W(t)} & f (\mathbf{k}) \ket{W (t)}
 = 2 \Im \varepsilon_{\mathbf{k}_c} - e \mathbf{E} \cdot \left[ \partial_\mathbf{k} f \right. \nonumber \\
 & \left. + f (\mathbf{k}_c) \left( \frac{\partial_\mathbf{k} I }{I} + i (\mathcal{A}^{LR} - \mathcal{A}^{RL}) \right) \right] |_{\mathbf{k}_c} ,
\end{align}
as quoted in the main text, Eq.~\eqref{eq:EOM1}.

\subsection{Time evolution of average position}

The derivation of the time evolution of the central position is straightforward, but requires some additional integral identities.
We start by noting that the average position
\begin{align}
 \mathbf{r}_c
 &= \frac{1}{N} \bra{W(t)}\mathbf{r}\ket{W(t)}
  = \frac{1}{N} \int_{\mathbf{k},{\mathbf{k}^\prime}} w_{\mathbf{k}}^* w_{\mathbf{k}^\prime} \bra{\psi^R_{\mathbf{k}}} \mathbf{r} \ket{\psi^R_{\mathbf{k}}} \nonumber \\
 &= \frac{i}{N} \int_{\mathbf{k}} \left[ |w_\mathbf{k}|^2 \braket{u^R_\mathbf{k}}{\partial_\mathbf{k} u_\mathbf{k}^R}
 + I (\mathbf{k}) w_\mathbf{k}^* \partial_\mathbf{k} w_\mathbf{k} \right]
\end{align}
can be expressed in terms of previously introduced quantities. To evaluate the second integral, we define the angle $\varphi$ via $w_\mathbf{k} = |w_\mathbf{k}| e^{i \varphi}$ and use
\begin{align}
 \frac{1}{N} \int_\mathbf{k} & f I w_\mathbf{k}^* \partial_j w_\mathbf{k}
 = \frac{1}{N} \int_\mathbf{k} f  I \left[ i |w_\mathbf{k}|^2 \partial_j \varphi + r \partial_j r \right] \nonumber \\
 &= i f \partial_j \varphi + \frac{1}{2N} \int_\mathbf{k} f I \partial_j |w_\mathbf{k}|^2 \nonumber \\
 &= f \left[ i \partial_j \varphi - \frac{1}{2} \frac{\partial_j I}{I} \right] - \frac{1}{2} \partial_j f , \label{eq:momentum_der_integral}
\end{align}
where we dropped all momentum labels whenever there is no ambiguity, and denote momentum derivatives with respect to the $j$th component of $\mathbf{k}$ by $\partial_j$, as we will do in the remainder of this section.
(Derivatives with respect to ${\mathbf{k}^\prime}$ are marked as such.)
The above relation allows us to express the central position
\begin{align}
 \mathbf{r}_c
 &= \left. \mathcal{A}^{RR} - \frac{i}{2} \frac{\partial_\mathbf{k} I}{I} - \partial_\mathbf{k} \varphi \right|_{\mathbf{k}= \mathbf{k}_c},
 \label{eq:central_position_appendix}
\end{align}
as quoted in the main text, Eq.~\eqref{eq:central_position}.
Similarly to Eq.~\eqref{eq:momentum_der_integral}, we obtain
\begin{align}
 \frac{1}{N} \int_\mathbf{k} f  I w_\mathbf{k} \partial_j w_\mathbf{k}^*
 = f \left[ -i \partial_j \varphi - \frac{1}{2} \frac{\partial_j I}{I} \right] - \frac{1}{2} \partial_j f
\end{align}
for integrals containing the term $w_\mathbf{k} \partial_j w_\mathbf{k}^*$.

We are interested in the time evolution of $\mathbf{r}_c$,
\begin{align}
 \dot{\mathbf{r}}_c = \frac{1}{N} \frac{d}{dt} \bra{W(t)} \mathbf{r}\ket{W (t)} - \frac{\dot{m}}{m} \mathbf{r}_c
\end{align}
and write the weight rate $\dot{N}/N$ in the short form
\begin{align}
 \frac{\dot{N}}{N} = 2 \Im \varepsilon - e \sum_j E_j \eta_j ,& &
 \eta_j = \frac{\partial_j I}{I} + i (\mathcal{A}^{LR}_j - \mathcal{A}^{RL}_j ) \label{eq:eta_def} .
\end{align}
For now, we focus on the $i$th component of the time derivative of $\bra{W(t)} \mathbf{r}\ket{W (t)}$, which we split up into two parts:
\begin{align}
 \frac{1}{N} & \frac{d}{dt} \bra{W(t)} x_i \ket{W (t)} = \dot{\alpha}_i + \dot{\beta}_i ,
\end{align}
with
\begin{align}
 \dot{\alpha}_i &= \frac{i}{N} \int_\mathbf{k} I ( \dot{w}_\mathbf{k}^*  \partial_i w_\mathbf{k} + w_\mathbf{k}^*  \partial_i \dot{w}_\mathbf{k} ) , \label{eq:alpha_dot} \\
 \dot{\beta}_i &= \frac{i}{N} \int_\mathbf{k} (\dot{w}_\mathbf{k}^* w_\mathbf{k} + w_\mathbf{k}^* \dot{w}_\mathbf{k}) \braket{u_\mathbf{k}^R}{\partial_i u_\mathbf{k}^R} . \label{eq:beta_dot}
\end{align}
Both terms contain a contribution independent of the electric field and another contribution that enters as the scalar product with $\mathbf{E}$,
\begin{align}
  \dot{\alpha}_i = \dot{\alpha}_i^\varepsilon + e \sum_j \dot{\alpha}_{ij}^E E_j , & &
   \dot{\beta}_i =  \dot{\beta}_i^\varepsilon + e \sum_j  \dot{\beta}_{ij}^E E_j .
\end{align}

We start by evaluating $\dot{\alpha}_i^\varepsilon$ and $\dot{\beta}_i^\varepsilon$.
Inserting only the field-independent contribution to $\dot{w}_{\mathbf{k}}$ into Eq.~\eqref{eq:alpha_dot} and using Eq.~\eqref{eq:momentum_der_integral} to evaluate the integral gives
\begin{align}
 \dot{\alpha}_i^\varepsilon
 &= \frac{1}{N} \int_\mathbf{k} I [ ( \varepsilon_\mathbf{k} - \varepsilon_\mathbf{k}^* ) w_\mathbf{k}^* \partial_i w_\mathbf{k} + |w_\mathbf{k}|^2 \partial_i \varepsilon_\mathbf{k} ] \nonumber \\
 &= ( \varepsilon_{\mathbf{k}} - \varepsilon_{\mathbf{k}}^* ) \left[ i \partial_i \varphi - \frac{1}{2} \frac{\partial_i I}{I} \right] - \frac{1}{2} \partial_i ( \varepsilon_\mathbf{k} - \varepsilon_\mathbf{k}^* ) + \partial_i \varepsilon_\mathbf{k} \nonumber \\
 &= 2 \Im \varepsilon_{\mathbf{k}_c} \left[ -\partial_i \varphi - \frac{i}{2} \frac{\partial_i I}{I} \right] + \partial_i \Re \varepsilon_\mathbf{k} .
\end{align}
Similarly, the field-independent contribution to Eq.~\eqref{eq:beta_dot} is
\begin{align}
 \dot{\beta}_i^\varepsilon
 =& \frac{1}{N} \int_\mathbf{k} I |w_\mathbf{k}|^2 ( \varepsilon_\mathbf{k} - \varepsilon_\mathbf{k}^* ) \braket{u_\mathbf{k}^R}{\partial_i u_\mathbf{k}^R} = 2  \mathcal{A}^{RR}_i \Im \varepsilon_\mathbf{k} ,
\end{align}
such that
\begin{align}
 \dot{\alpha}_i^\varepsilon +  \dot{\beta}_i^\varepsilon
 =& 2 \Im \varepsilon_\mathbf{k} \left[\mathcal{A}_i^{RR} - \frac{i}{2} \frac{\partial_i I}{I}  - \partial_i \varphi \right] + \partial_i \Re \varepsilon_\mathbf{k} \nonumber \\
 =& 2 (r_c)_i \Im \varepsilon_\mathbf{k} + \partial_i \Re \varepsilon_\mathbf{k} ,
 \label{eq:alpha_beta_epsilon}
\end{align}
where we inserted the central position coordinate $(r_c)_i$ using Eq.~\eqref{eq:central_position_appendix}

The field-dependent contribution is more challenging to evaluate.
We start by noting that $\alpha_{ij}^E$ consists of three parts,
\begin{widetext}
\begin{align}
  \dot{\alpha}_{ij}^E
  =&  \frac{1}{N} \int_{\mathbf{k},{\mathbf{k}^\prime}} I (\mathbf{k})
   \left[ w_\mathbf{k}^* w_{{\mathbf{k}^\prime}} \bra{\partial_i \psi_\mathbf{k}^L} x_j \ket{\psi_{{\mathbf{k}^\prime}}^R}  - w_{{\mathbf{k}^\prime}}^*\bra{\psi_{{\mathbf{k}^\prime}}^R} x_j \ket{\psi_{\mathbf{k}}^L} \partial_i w_\mathbf{k} \right] \\
  =&  - \frac{1}{N} \int_{\mathbf{k},{\mathbf{k}^\prime}}
   \left[
      w_{{\mathbf{k}^\prime}} \partial_i ( I(\mathbf{k}) w_\mathbf{k}^* ) \bra{\psi_{\mathbf{k}}^L} x_j \ket{\psi_{{\mathbf{k}^\prime}}^R} + I(\mathbf{k}) w_{{\mathbf{k}^\prime}}^* \partial_i w_\mathbf{k} \bra{\psi_{{\mathbf{k}^\prime}}^R} x_j \ket{\psi_{\mathbf{k}}^L} \right],
\end{align}
containing derivatives $\partial_i$ with respect to $w_\mathbf{k}^*$, $I(\mathbf{k})$, and $w_\mathbf{k}$, respectively.
We simplify each of these terms one by one.
Using Eq.~\eqref{eq:r_expectation} gives
\begin{align}
  \dot{\alpha}_{ij}^E
  =& - \frac{i}{N} \int_\mathbf{k} I  \left[ w_\mathbf{k} \partial_i w_\mathbf{k}^* \braket{u_\mathbf{k}^L}{\partial_j u_\mathbf{k}^R} + \partial_j w_\mathbf{k} \partial_i w_\mathbf{k}^*
 + w_\mathbf{k}^* \partial_i w_\mathbf{k} \left( \braket{u_\mathbf{k}^R}{\partial_j u_\mathbf{k}^L}  + \frac{\partial_j I}{I} \right) + w_\mathbf{k}^* \partial_j \partial_i w_\mathbf{k} \right] \\
   &- \frac{i}{N} \int_{\mathbf{k}} \partial_i I \left[ |w_\mathbf{k}|^2  \braket{u_\mathbf{k}^L}{\partial_j u_\mathbf{k}^R} + w_\mathbf{k}^* \partial_j w_\mathbf{k} \right] \nonumber .
\end{align}
The first integral that contains $w_\mathbf{k} \partial_i w_\mathbf{k}^*$ and $w_\mathbf{k}^* \partial_i w_\mathbf{k}$ can be solved using Eq.~\eqref{eq:momentum_der_integral}.
Further simplifying the first term in the second integral gives
\begin{align}
  \dot{\alpha}_{ij}^E
  =& -\left( \frac{\partial_j I}{I} + i (\mathcal{A}^{LR}_j - \mathcal{A}^{RL}_j) \right) \left( -\partial_i \varphi - i \frac{\partial_i I}{2 I} \right) + \frac{1}{2} \partial_i \left( \mathcal{A}^{LR}_j +  \mathcal{A}^{RL}_j \right) + \frac{i}{2} \partial_i \frac{\partial_j I}{I} \nonumber \\
  &-  \frac{i}{N} \int_\mathbf{k} \left[ I w_\mathbf{k}^* \partial_j \partial_i w_\mathbf{k} + \partial_i I w_\mathbf{k}^* \partial_j w_\mathbf{k} +  I \partial_i w_\mathbf{k}^* \partial_j w_\mathbf{k} \right] .
\end{align}
The last remaining integral is zero,
\begin{equation}
\int_\mathbf{k} \left[ I  w_\mathbf{k}^* \partial_i \partial_j w_\mathbf{k} + \partial_i I w_\mathbf{k}^* \partial_j w_\mathbf{k} + I  \partial_i w_\mathbf{k}^* \partial_j w_\mathbf{k}\right] = \int_\mathbf{k} \partial_i (  I  w_\mathbf{k}^* \partial_j w_\mathbf{k} ) = 0 ,
\end{equation}
such that we obtain the simplified form
\begin{align}
 \dot{\alpha}_{ij}^E
  =& - \left( \frac{\partial_j I}{I} +  i (\mathcal{A}^{LR}_j - \mathcal{A}^{RL}_j) \right) \left( - \partial_i \varphi - i \frac{\partial_i I}{2I} \right) + \frac{1}{2} \partial_i \left( \mathcal{A}^{LR}_j + \mathcal{A}^{RL}_j \right)
  + i \partial_i \frac{\partial_j I}{2 I} \nonumber \\
  =& - \eta_j \left[ -\partial_i \varphi - \frac{i}{2} \frac{\partial_i I}{I} \right] 
  + \partial_i \frac{ \mathcal{A}^{LR}_j + \mathcal{A}^{RL}_j }{2}
  + i \partial_j \frac{\partial_i I}{2 I} ,
\end{align}
where we used that $\partial_i ((\partial_j I)/I) = \partial_j ((\partial_i I)/I)$ and introduced $\eta_j$ from Eq.~\eqref{eq:eta_def}.
The contribution $\dot{\beta}_{ij}^E$ is easier to evaluate. We first use Eq.~\eqref{eq:r_expectation} to simplify
\begin{align}
 \dot{\beta}_{ij}^E
 =&	\frac{1}{N} \int_{\mathbf{k},{\mathbf{k}^\prime}} [ w_\mathbf{k}^* w_{\mathbf{k}^\prime} \bra{\psi_\mathbf{k}^L} x_j \ket{\psi_{{\mathbf{k}^\prime}}^R} - w_{\mathbf{k}^\prime}^* w_\mathbf{k} \bra{\psi_{{\mathbf{k}^\prime}}^R} x_j \ket{\psi_\mathbf{k}^L} ] \braket{u_\mathbf{k}^R}{\partial_i u_\mathbf{k}^R} \\
 =& \frac{i}{N} \int_\mathbf{k} |w_\mathbf{k}|^2 [\braket{u_\mathbf{k}^L}{\partial_j u_\mathbf{k}^R} - \braket{u_\mathbf{k}^R}{\partial_j u_\mathbf{k}^L} ] \braket{u_\mathbf{k}^R}{\partial_i u_\mathbf{k}^R} - \frac{i}{N} \int_\mathbf{k} |w_\mathbf{k}|^2 \partial_j \braket{u_\mathbf{k}^R}{\partial_i u_\mathbf{k}^R} ,
\end{align}
which gives
\begin{equation}
 \dot{\beta}_{ij}^E
 = - \left[  \frac{\partial_j I}{I} + i(\mathcal{A}_j^{LR} - \mathcal{A}_j^{RL}) \right] \mathcal{A}_i^{RR} - \partial_j  \mathcal{A}_i^{RR} = -\eta_j \mathcal{A}_i^{RR} - \partial_j \mathcal{A}_i^{RR} .
\end{equation}
\end{widetext}
The sum of the two field-dependent contributions is therefore
\begin{align}
 \dot{\alpha}_{ij}^E + \dot{\beta}_{ij}^E
 =& \partial_i \frac{ \mathcal{A}^{LR}_j + \mathcal{A}^{RL}_j}{2} - \partial_j \left( \mathcal{A}_i^{RR} - \frac{i}{2} \frac{\partial_i I}{I} \right) - \eta_j x_i ,
\end{align}
which, combined with Eq.~\eqref{eq:alpha_beta_epsilon}, gives
\begin{align}
 \frac{1}{N} & \frac{d}{dt} \bra{W} x_i \ket{W}
  = (r_c)_i (2 \Im \varepsilon_\mathbf{k} - e E_j \eta_j ) + \partial_i \Re \varepsilon_\mathbf{k} \nonumber  \\
  &+ e E_j \left[ \partial_i \frac{ \mathcal{A}^{LR}_j + \mathcal{A}^{RL}_j}{2} - \partial_j \left( \mathcal{A}_i^{RR} - \frac{i}{2} \frac{\partial_i I}{I} \right)  \right] .
\end{align}
We finally identify $\dot{N}/N = 2 \Im \varepsilon_\mathbf{k} - e E_j \eta_j$ to write the time evolution of the central position as
\begin{align}
 (\dot{r}_c)_i = \partial_i \Re \varepsilon_\mathbf{k} - e E_j
  & \left[  \partial_j \left( \mathcal{A}_i^{RR} - \frac{i}{2} \frac{\partial_i I}{I} \right) \right. \nonumber \\
  & \left. - \frac{1}{2} \partial_i ( \mathcal{A}^{LR}_j + \mathcal{A}^{RL}_j ) \right]
\end{align}
as quoted in the main text, Eq.~\eqref{eq:rdot}.

\section{Gauge invariance}
\label{sec:gauge_invariance}

In this appendix, we show that the anomalous weight rate and velocity are independent of the gauge choice.
Here, we understand ``gauge choice'' as a generalization of the gauge in Hermitian systems.
In particular, a transformation of the right cell-periodic eigenstates
\begin{equation}
 \ket{u_{n \mathbf{k}}^R} \to \ket{\bar{u}_{n \mathbf{k}}^R} = f_n (\mathbf{k}) \ket{u_{n \mathbf{k}}^R}
\end{equation}
with an analytic and nonzero function $f_n (\mathbf{k})$ requires a corresponding transformation of the left eigenstates
\begin{equation}
 \ket{u_{n \mathbf{k}}^L} \to \ket{\bar{u}_{n \mathbf{k}}^L} = \frac{1}{f_n^* (\mathbf{k})} \ket{u_{n \mathbf{k}}^L}
\end{equation}
to keep left and right eigenstates orthonormal~\cite{Shen:2018bx}.
This implies that
\begin{align}
 \bar{I}_{nn^\prime } (\mathbf{k})
 &= \braket{ \bar{u}_{n \mathbf{k}}^R}{ \bar{u}_{n^\prime  \mathbf{k}}^R }
  = f_n^* (\mathbf{k}) f_n (\mathbf{k}) I_{nn^\prime } (\mathbf{k})
\end{align}
and
\begin{align}
 \frac{\partial_\mathbf{k} \bar{I}_{nn} (\mathbf{k})}{ \bar{I}_{nn} (\mathbf{k})}
 =  \frac{\partial_\mathbf{k} I_{nn} (\mathbf{k})}{I_{nn} (\mathbf{k})}
 + \frac{\partial_\mathbf{k} f_n^*(\mathbf{k})}{f_n^*(\mathbf{k})}
 + \frac{\partial_\mathbf{k} f_n  (\mathbf{k})}{f_n  (\mathbf{k})} .
\end{align}
The Berry connections transform accordingly
\begin{align}
 \bar{\mathcal{A}}_n^{LR} = \mathcal{A}_n^{LR} + i \frac{ \partial_\mathbf{k} f  (\mathbf{k})}{f_n  (\mathbf{k})} , & &
 \bar{\mathcal{A}}_n^{RL} = \mathcal{A}_n^{RL} - i \frac{ \partial_\mathbf{k} f^*(\mathbf{k})}{f_n^*(\mathbf{k})} ,
 \label{eq:gauge_berry_LRRL}
\end{align}
which gives
\begin{align}
 \frac{\partial_\mathbf{k} \bar{I} (\mathbf{k})}{\bar{I} (\mathbf{k})}
   + i ( \bar{\mathcal{A}}^{LR} -  \bar{\mathcal{A}}^{RL} ) 
 =  \frac{\partial_\mathbf{k} I (\mathbf{k})}{I (\mathbf{k})} + i ( \mathcal{A}^{LR} - \mathcal{A}^{RL} ) ,
\end{align}
i.e., the anomalous weight rate is gauge-invariant.

To confirm that the anomalous velocity is gauge-independent, we start by noting that $\mathcal{A}_n^{RR}$ transforms as
\begin{align}
 \bar{\mathcal{A}}_n^{RR} =  \mathcal{A}_n^{RR} + \frac{i \partial_\mathbf{k} f_n (\mathbf{k})}{f_n (\mathbf{k})}
\end{align}
and accordingly
\begin{align}
 \bar{\mathcal{A}}_n^{RR} - \frac{i}{2}  \frac{\partial_\mathbf{k} \bar{I}_{nn} (\mathbf{k})}{\bar{I}_{nn} (\mathbf{k})}
 =&  \mathcal{A}_n^{RR}  - \frac{i}{2} \frac{\partial_\mathbf{k} I_{nn} (\mathbf{k})}{I_{nn} (\mathbf{k})} \\
  &+ \frac{i}{2} \left(
    \frac{\partial_\mathbf{k} f_n (\mathbf{k})}{f_n (\mathbf{k})}
 -  \frac{\partial_\mathbf{k} f_n^*(\mathbf{k})}{f_n^*(\mathbf{k})} \right) \nonumber .
\end{align}
Using Eq.~\eqref{eq:gauge_berry_LRRL} for the Berry connection with $\alpha\neq\beta$, we obtain
\begin{align}
 \bar{\mathcal{A}}_n^{LR} +  \bar{\mathcal{A}}_n^{RL}
 &= \mathcal{A}_n^{LR} + \mathcal{A}_n^{RL}
 + i \frac{ \partial_\mathbf{k} f_n  (\mathbf{k})}{f_n  (\mathbf{k})}
 - i \frac{ \partial_\mathbf{k} f_n^*(\mathbf{k})}{f_n^*(\mathbf{k})} ,
\end{align}
such that, using the symmetry of second derivatives,
\begin{equation}
 \partial_j \left[ \frac{\partial_i f_n (\mathbf{k})}{f_n (\mathbf{k})} \right] -  \partial_i \left[ \frac{\partial_j f_n (\mathbf{k})}{f_n (\mathbf{k})} \right]
 = 0 ,
\end{equation}
the terms containing $f_n$ cancel in the anomalous velocity term.
Thus, both quantities are gauge-independent.

\section{Computing the anomalous weight rate and velocity using projectors}
\label{sec:projectors}

When numerically finding the eigenstates of a finite-dimensional Hamiltonian at two different momenta, the phase of these two eigenstates will be essentially independent of each other. In other words, it is hard to fix a gauge for the whole Brillouin zone numerically. This implies that a naive computation of the momentum derivative of an eigenstates $\partial_\mathbf{k} \ket{u_{n\mathbf{k}}^\alpha}$ via finite differences cannot be used due to the independent phases at $\mathbf{k}$ and $\mathbf{k} + d\mathbf{k}$. Thus, to compute the anomalous weight rate and velocity numerically, we need a method that is independent of the (random) phase of the eigenstates.
Using a biorthogonal basis [Eq.~\eqref{eq:biorthogonal_states}], the projector
\begin{align}
 P^{RL} = \ket{u_{n\mathbf{k}}^R} \bra{u_{n\mathbf{k}}^L}
\end{align}
and its Hermitian conjugate $(P^{RL})^\dagger =  P^{LR}$ are invariant under the transformation $\ket{u_{n\mathbf{k}}^R} \to f_n(\mathbf{k}) \ket{u_{n\mathbf{k}}^R}$ with an analytic and nonzero function $f_n (\mathbf{k})$, as long as the orthonormality condition~\eqref{eq:biorthogonal_states} is satisfied. By further normalizing the right eigenstates according to $I_{nn} (\mathbf{k}) = 1$,  all projectors $P^{\alpha\beta} = \ket{u_{n\mathbf{k}}^\alpha} \bra{u_{n\mathbf{k}}^\beta}$ with $\alpha,\beta = L,R$ are gauge-independent.

Thus, using the normalization $\braket{u_{n\mathbf{k}}^\alpha}{u_{n\mathbf{k}}^R} = 1$, we find that the trace over the band indices at fixed $\mathbf{k}$
\begin{align}
 \tr & [ P^{RR} \partial_\mathbf{k} P^{LL} ]
 = \tr \left[ \ket{\partial_\mathbf{k} u_{n\mathbf{k}}^L} \bra{u_{n\mathbf{k}}^R} \right]
 + \tr \left[ \ket{u_{n\mathbf{k}}^R} \bra{\partial_\mathbf{k} u_{n\mathbf{k}}^L} \right] \nonumber \\
 &= \braket{u_{n\mathbf{k}}^R}{\partial_\mathbf{k} u_{n\mathbf{k}}^L} + \braket{\partial_\mathbf{k} u_{n\mathbf{k}}^L}{u_{n\mathbf{k}}^R}
 = i \left( \mathcal{A}^{LR} - \mathcal{A}^{RL} \right) .
\end{align}
Since in our gauge $\partial_\mathbf{k} I_{nn} (\mathbf{k})=0$, the above expression equals the Berry-connection-induced contribution to the anomalous weight rate, Eq.~\eqref{eq:dmdt}.

In 1D systems, we also find an expression for the anomalous velocity contribution in terms of gauge-invariant quantities. We use the relations
\begin{align}
 \tr \left[ P^{LR} \partial_k P^{RR} \right]
  =& \tr \left[ \ket{\partial_k u_{nk}^R} \bra{u_{nk}^R}  + \ket{u_{nk}^L} \bra{\partial_k u_{nk}^R} \right] \nonumber \\
  =& \braket{u_{nk}^R}{\partial_k u_{nk}^R} + \braket{\partial_k u_{nk}^R}{u_{nk}^L}
\end{align}
and
\begin{align}
 \tr \left[ P^{RL} \partial_k P^{RR} \right]
  =& \tr \left[  \ket{\partial_k u_{nk}^R} \bra{u_{nk}^L} + \ket{u_{nk}^R} \bra{\partial_k u_{nk}^R} \right] \nonumber \\
  =& \braket{u_{nk}^L}{\partial_k u_{nk}^R} + \braket{\partial_k u_{nk}^R}{u_{nk}^R} .
\end{align}
We identify the different Berry connections and again use $I_{nn} (\mathbf{k})=1$, which gives
\begin{equation}
 \frac{i}{2} \tr \left[ (P^{RL} - P^{LR}) \partial_k P^{RR} \right]
 = \frac{1}{2} ( \mathcal{A}^{LR} + \mathcal{A}^{RL} ) - \mathcal{A}^{RR} .
\end{equation}
Taking the derivative with respect to $k$ gives the Berry-connection-induced contribution to the anomalous velocity, Eq.~\eqref{eq:rdot1D}.
A similar approach applies to higher dimension, although one would have to start from combinations of expressions of the form $\tr [ \partial_{k_i} P^{\alpha\beta} \partial_{k_j} P^{\alpha^\prime\beta^\prime} ]$ to obtain the combination of Berry connection derivatives appearing in Eq.~\eqref{eq:rdot}.

\section{Symmetry restrictions on anomalous weight rate and velocity}
\label{sec:appendix_symmetries}

Similar to Hermitian systems where the anomalous velocity term is zero when both time-reversal and inversion symmetry are present~\cite{Xiao:2010kw}, symmetries play a central role in determining the anomalous velocity contribution in non-Hermitian systems.
In this appendix, we discuss in detail how antiunitary symmetries and their combinations, as well as inversion symmetry, constrain the anomalous weight rate and velocity terms.

In non-Hermitian systems, time-reversal and particle-hole symmetry formally get promoted to four different antiunitary symmetries~\cite{Kawabata:2019en}:
\begin{align}
 T_\pm H^*_{-\mathbf{k}} T_\pm^\dagger &= \pm H_{\mathbf{k}} , \label{eq:Tplusminus} \\
 C_\pm H^T_{-\mathbf{k}} C_\pm^\dagger &= \pm H_{\mathbf{k}} , \label{eq:Cplusminus}
\end{align}
where all matrices $T_\pm$ and $C_\pm$ are unitary.
We do not aim to discuss the physical meaning of these operations, but only investigate the effect of these constraints on the anomalous terms in the semiclassical equations of motion.
The constraints involving $T_+$ and $C_+$ commute with the Hamiltonian and can thus be understood as time-reversal symmetries, whereas and the constraints involving $T_-$ and $C_-$ anticommute with the Hamiltonian and can be understood as particle-hole symmetries.
Similar to Hermitian Hamiltonians, where chiral symmetry, the combination of time-reversal and particle-hole symmetry, may still be present when both symmetries are broken individually~\cite{Altland:1997cg}, the combinations of the antiunitary symmetries considered above are also relevant for the symmetry classification.
Following Ref.~\onlinecite{Kawabata:2019en}, we identify the sublattice symmetry $S = T_+ T_-^*$ (or $S=C_+ C_-^T$), chiral symmetry $\Gamma = T_+ C_-^*$ (or $\Gamma = T_- C_+^*$) and pseudo-Hermiticity $\eta = T_+ C_+^*$ (or $\eta = T_- C_-^*$), giving
\begin{align}
 \mathcal{S} H_{\mathbf{k}} \mathcal{S}^\dagger &= -H_{\mathbf{k}} \\
 \Gamma H_{\mathbf{k}}^\dagger \Gamma^\dagger &= -H_{\mathbf{k}} \\
 \eta H^\dagger_{\mathbf{k}} \eta^\dagger &= H_{\mathbf{k}} \label{eq:pseudo_hermiticity},
\end{align}
with unitary $\mathcal{S}$, $\Gamma$, and $\eta$.
Since it is plays a key role in non-Hermitian systems~\cite{Bender:1998bw}, we also consider inversion symmetry,
\begin{align}
 P H_{-\mathbf{k}} P^\dagger &= H_{\mathbf{k}},
\end{align}
with unitary $P$. Differently from the other symmetries considered above, it is a unitary symmetry that commutes with the Hamiltonian.

We focus on time-reversal symmetry, pseudo-Hermiticity, and inversion symmetry. Different from Hermitian systems, particle-hole symmetries might also play a crucial role in determining restrictions on the anomalous weight rate and velocity, because they can restrict energies to be purely imaginary~\cite{Kawabata:2019fd}.
Since it is beyond the scope of this paper, we leave the investigation of particle-hole symmetries for future work.

In the following, we do not use bra-ket notation, but instead consider a finite-dimensional Hilbert space where a matrix $U_\mathbf{k}$ that diagonalizes the tight-binding Hamiltonian exists, which implies the absence of exceptional points for all $\mathbf{k}$~\footnote{Since the eigenvectors at exceptional points coalesce, $\det U_\mathbf{k}=0$ and $U_\mathbf{k}$ is not invertible.}.
Using $U_\mathbf{k}$, the diagonal form of the Hamiltonian reads
\begin{align}
 \diag (\{\varepsilon_{n,\mathbf{k}} \})    &= U^{-1}_{\mathbf{k}} H_{\mathbf{k}} U_{\mathbf{k}} ,
 \label{eq:diagonalization_ham}
\end{align}
where generally $U_{\mathbf{k}}^{-1} \neq U_{\mathbf{k}}^\dagger$.
For simplicity, we choose the Hamiltonian such that its eigenstates are the lattice-periodic part of the eigenvectors~\cite{Alexandradinata:2016kb}.
The columns of $U_\mathbf{k}$ are the right eigenvectors of $\mathcal{H}_\mathbf{k}$ and the rows of $(U_\mathbf{k}^{-1})^*$ the left eigenvectors, in other words,
\begin{align}
 [ U_{\mathbf{k}} ]_{jn} = \braket{j}{u_{n\mathbf{k}}^R} , & & 
 [U_{\mathbf{k}}^{-1}]_{nj} = \braket{u_{n\mathbf{k}}^L}{j} .
\end{align}
We first consider the two variants of time-reversal symmetry before turning to their combination with inversion symmetry and finally to pseudo-Hermiticity.

\subsection{Time-reversal symmetry $T_+$}

To investigate the effect of time-reversal, we first take the complex conjugate of Eq.~\eqref{eq:diagonalization_ham} and then insert Eq.~\eqref{eq:Tplusminus}
\begin{align}
 \diag (\{\varepsilon_{n,-\mathbf{k}}^* \})
 &= (U_{-\mathbf{k}}^*)^{-1} H_{-\mathbf{k}}^* U_{-\mathbf{k}}^* \nonumber \\
 &= ( T_+ U^*_{-\mathbf{k}} )^{-1} H_{\mathbf{k}} T_+ U_{-\mathbf{k}}^* \label{eq:restrictions_Tplus}.
\end{align}
This implies that for each state with energy $\varepsilon_{n,\mathbf{k}}$ at $\mathbf{k}$, there must exist a state with energy $\varepsilon_{n^\prime ,-\mathbf{k}} = \varepsilon_{n,\mathbf{k}}^*$ at $-\mathbf{k}$.
At time-reversal invariant momenta (TRIM) $\mathbf{K}$, two linearly independent states with different energies $\varepsilon_{n,\mathbf{K}} \neq \varepsilon_{n^\prime ,\mathbf{K}}$ (yet related via complex conjugation $\varepsilon_{n,\mathbf{K}} = \varepsilon_{n^\prime ,\mathbf{K}}^*$) must exist when $T_+ T_+^* = -1$. Differently, time-reversal symmetry with $T_+ T_+^* = +1$ relates the state to itself at $\mathbf{K}$, which results in $\varepsilon_{n,\mathbf{K}} = \varepsilon_{n,\mathbf{K}}^*$, i.e., the energies are purely real at TRIM.
From Eq.~\eqref{eq:restrictions_Tplus}, one is tempted to conclude that
\begin{equation}
 U_\mathbf{k} = T_+ U_{-\mathbf{k}}^* \label{eq:tr_plus_phase}.
\end{equation}
However, this is only a sensible choice when $T_+ T_+^* = +1$. Otherwise, the energy bands $\varepsilon_{n,\mathbf{k}}$ would be discontinuous at the TRIM.

Time-reversal symmetry $T_+ T_+^* = +1$ has several implications for the anomalous velocity and weight rate terms.
Fixing the phase relation via Eq.~\eqref{eq:tr_plus_phase}, the Gramian matrix of the right eigenstates is constrained by
\begin{align}
 I_{nn^\prime } (\mathbf{k})
 &= \sum_j [U_{\mathbf{k}}^*]_{jn} [U_{\mathbf{k}}]_{jn^\prime } \nonumber \\
 &= \sum_j [ T_+^* U_{-\mathbf{k}}]_{jn} [T_+ U_{-\mathbf{k}}^* ]_{jn^\prime } = I_{n^\prime n} (-\mathbf{k}) .
\end{align}
Constraints set by $T_+$ also apply to the four non-Hermitian generalizations of the Berry connection, Eq.~\eqref{eq:berry_connection}.
Using the matrix $U_\mathbf{k}$, the terms $\mathcal{A}_n^{\alpha\beta}$ with $\alpha \neq \beta$ read
\begin{subequations}\begin{align}
 \mathcal{A}^{LR}_n (\mathbf{k})
  &= i \sum_j [U^{-1}_{\mathbf{k}}]_{nj} \partial_\mathbf{k} [U_{\mathbf{k}}]_{jn},  \\
 \mathcal{A}^{RL}_n (\mathbf{k})
 &=i \sum_j [U_{\mathbf{k}}^{*}]_{jn} \partial_\mathbf{k} [U_{\mathbf{k}}^{-1}]^*_{nj} ,
\end{align}\label{eq:berry_matrix_form}\end{subequations}
which gives, employing Eq.~\eqref{eq:tr_plus_phase}
\begin{align}
 \mathcal{A}^{LR}_n (\mathbf{k})
  &= i \sum_j [U^{-1}_{-\mathbf{k}}]^*_{nj} \partial_\mathbf{k} [U^*_{-\mathbf{k}}]_{jn} \label{eq:tplus_LR} \\
  &=-i \sum_j [U_{-\mathbf{k}}^*]_{jn} \partial_\mathbf{k} [U^{-1}_{-\mathbf{k}}]^*_{nj} 
   = -\mathcal{A}^{RL}_n (-\mathbf{k}) \nonumber,
\end{align}
where we used that $\partial_\mathbf{k} (U_{\mathbf{k}}^{-1} U_\mathbf{k}) = 0$.
Since the two Berry connections are further related via $[\mathcal{A}^{RL}_n (-\mathbf{k})]^* = \mathcal{A}^{LR}_n (-\mathbf{k})$, integrals of $\mathcal{A}^{RL}_n$ and $\mathcal{A}^{LR}_n$ over the whole BZ are purely imaginary in presence of $T_+$.
Furthermore, since the derivative $\partial_\mathbf{k} I_{nn} (\mathbf{k})$ of the even function $I_{nn} (\mathbf{k})$ is odd, the anomalous weight rate integrated over the whole Brillouin zone vanishes.

The Berry connection $\mathcal{A}_n^{\alpha\alpha}$ at $\mathbf{k}$ is related to itself at $-\mathbf{k}$ via
\begin{align}
 \mathcal{A}^{RR}_n (\mathbf{k})
  &= i \frac{\sum_j [ T_+^* U_{-\mathbf{k}}]_{jn} \partial_\mathbf{k} [T_+ U_{-\mathbf{k}}^*]_{jn}}{I_{nn} (\mathbf{k})} \nonumber \\
  &= -i \frac{\sum_j [ U_{-\mathbf{k}}^*]_{jn} \partial_\mathbf{k} [U_{-\mathbf{k}}]_{jn} - \partial_\mathbf{k} I_{nn} (-\mathbf{k})}{I_{nn} (\mathbf{k})} \nonumber \\
  &= - \mathcal{A}^{RR}_{n} (-\mathbf{k}) + i  \frac{\partial_\mathbf{k} I_{nn} (-\mathbf{k})}{I_{nn} (-\mathbf{k})} , \label{eq:tplus_RR}
\end{align}
and
\begin{align}
 \mathcal{A}^{LL}_n (\mathbf{k}) 
  &= i\frac{\sum_j [ (T_+ U_{-\mathbf{k}}^*)^{-1}]_{nj} \partial_\mathbf{k} [ (T_+^* U_{-\mathbf{k}})^{-1}]_{nj}}{[I^{-1}]_{nn} (-\mathbf{k})} \nonumber \\
  &= -i\frac{\sum_j [ U_{-\mathbf{k}}^{-1} ]_{nj} \partial_\mathbf{k} [ U_{-\mathbf{k}}^{-1}]^*_{nj} - \partial_\mathbf{k} [I^{-1}]_{nn} (-\mathbf{k})}{[I^{-1}]_{nn} (-\mathbf{k})} \nonumber  \\
  &= - \mathcal{A}^{LL}_{n} (-\mathbf{k}) + i \frac{\partial_\mathbf{k} [ I^{-1} ]_{nn} (-\mathbf{k})}{[ I^{-1} ]_{nn}  (-\mathbf{k})} .
\end{align}
Using Eq.~\eqref{eq:reality_conditions}, we conclude that $\Re \mathcal{A}_n^{RR} (\mathbf{k}) = - \Re \mathcal{A}_n^{RR} (-\mathbf{k})$ and $\Re \mathcal{A}_n^{LL} (\mathbf{k}) = - \Re \mathcal{A}_n^{LL} (-\mathbf{k})$.
The integrals of $\Re \mathcal{A}^{RR}_n (\mathbf{k})$ and $\Re \mathcal{A}^{LL}_n (\mathbf{k})$ over the whole BZ are therefore zero.

\subsection{Time-reversal symmetry $C_+$}

Time-reversal symmetry $C_+$ gives rise to different constraints on the Berry connections than $T_+$.
To fix the gauge of the eigenstates, we first take the transpose of Eq.~\eqref{eq:diagonalization_ham} and then insert Eq.~\eqref{eq:Cplusminus}
\begin{align}
 \diag (\{\varepsilon_{n,-\mathbf{k}} \})
 &= U_{-\mathbf{k}}^T H_{-\mathbf{k}}^T (U_{-\mathbf{k}}^T )^{-1} \\
 &= U_{-\mathbf{k}}^T C_+^\dagger H_{\mathbf{k}} ( U_{-\mathbf{k}}^T C_+^\dagger)^{-1} \label{eq:restrictions_Cplus}.
\end{align}
For each state with energy $\varepsilon_{n,\mathbf{k}}$ at $\mathbf{k}$, a state with the same energy $\varepsilon_{n^\prime ,-\mathbf{k}} = \varepsilon_{n,\mathbf{k}}$ must exist at $-\mathbf{k}$.
At TRIM $\mathbf{K}$, two degenerate states with energy $\varepsilon_{n,\mathbf{K}}$ must exist when $C_+ C_+^* = -1$, whereas time-reversed partners at $\mathbf{K}$ only differ by a phase when $C_+ C_+^* = +1$.
Different from $T_+$, we do not obtain a reality condition at $\mathbf{K}$.
We fix the phase relation between time-reversed partners at $\mathbf{k} \neq \mathbf{K}$ via
\begin{equation}
 U_\mathbf{k}^{-1} = U_{-\mathbf{k}}^T C_+^\dagger \label{eq:cr_plus_phase}.
\end{equation}
When $C_+ C_+^* =-1$, this definition needs to be slightly modified to be valid, e.g., by restricting $\mathbf{k}$ to half of the BZ. Here, we focus on $C_+ C_+^*=+1$ to avoid these subtleties.

When fixing the gauge according to Eq.~\eqref{eq:cr_plus_phase}, the Gramian matrix is related to its inverse via
\begin{align}
 I_{nn^\prime } (\mathbf{k})
 &= \sum_j [U_{\mathbf{k}}^*]_{jn} [U_{\mathbf{k}}]_{jn^\prime } \nonumber \\ 
 &= \sum_j [ C_+^* (U_{-\mathbf{k}}^{-1})^\dagger ]_{jn} [C_+ (U_{-\mathbf{k}}^{-1})^T ]_{jn^\prime } \nonumber \\ 
 &= [I^{-1}]_{n^\prime n} (-\mathbf{k}) \label{eq:cplus_overlap}.
\end{align}
The Berry connection $\mathcal{A}_n^{\alpha\beta}$ with $\alpha \neq \beta$ [Eq.~\eqref{eq:berry_matrix_form}] is
\begin{align}
 \mathcal{A}^{LR}_n (\mathbf{k})  \label{eq:cplus_LR}
  &= i \sum_j [U_{-\mathbf{k}}]_{jn} \partial_\mathbf{k} [U_{-\mathbf{k}}^{-1}]_{nj}\\
  &=-i \sum_j [U_{-\mathbf{k}}^{-1}]_{nj} \partial_\mathbf{k} [U_{-\mathbf{k}}]_{jn} = -\mathcal{A}^{LR}_n (-\mathbf{k}),  \nonumber
\end{align}
and analogously $\mathcal{A}^{RL}_n (\mathbf{k}) = - \mathcal{A}^{RL}_n (-\mathbf{k})$.
The Berry connection $\mathcal{A}_n^{\alpha\alpha}$ is constrained by
\begin{align}
 \mathcal{A}^{RR}_n (\mathbf{k})
  &= i \frac{\sum_j [ U_{-\mathbf{k}}^{-1} C_+]^*_{nj} \partial_\mathbf{k} [ U_{-\mathbf{k}}^{-1} C_+]_{nj}}{[I^{-1}]_{nn} (-\mathbf{k})} \nonumber \\
  &= -i \frac{\sum_j [ U_{-\mathbf{k}}^{-1}]_{nj} \partial_\mathbf{k} [ U_{-\mathbf{k}}^{-1}]^*_{nj} - \partial_\mathbf{k} [I^{-1}]_{nn} (-\mathbf{k})}{[I^{-1}]_{nn} (-\mathbf{k})} \nonumber \\
  &= - \mathcal{A}^{LL}_{n} (-\mathbf{k}) + i  \frac{\partial_\mathbf{k} [I^{-1}]_{nn} (-\mathbf{k})}{[I^{-1}]_{nn} (-\mathbf{k})} ,
  \label{eq:cplus_RR}
\end{align}
which implies $\Re \mathcal{A}_n^{RR} (\mathbf{k}) = - \mathcal{A}_n^{LL} (-\mathbf{k})$; cf.\ Eq.~\eqref{eq:reality_conditions}.
While the integrals of $\mathcal{A}^{RL}_n$ and $\mathcal{A}^{LR}_n$ over the whole BZ must be zero, the integral of $\Re \mathcal{A}^{RR}_n$ does not necessarily vanish, but must equal the integral of $\Re \mathcal{A}^{LL}_n$.

\subsection{Inversion and time-reversal $P T_+$}

The anomalous velocity term is zero in the presence of unbroken $PT_+$ symmetry with $(PT_+)(PT_+)^* =+1$.
To show that the anomalous velocity vanishes, we first need to fix the gauge of the eigenstates.
When it is possible to modify Eq.~\eqref{eq:tr_plus_phase} such that
\begin{equation}
 U_\mathbf{k} = P T_+ U_\mathbf{k}^* \label{eq:ptplus_gauge} ,
\end{equation}
all energies are real, and the $P T_+$ symmetry is said to be \emph{unbroken}~\cite{Bender:1998bw,Bender:2007kr}.
The unbroken $PT_+$ symmetry is separated by an exceptional point from the $PT_+$ broken phase, where eigenvalues are generally complex~\cite{Heiss:2012bx}.
Here, we consider the case when Eq.~\eqref{eq:ptplus_gauge} holds and we have purely real eigenvalues in a non-Hermitian system.
This implies that
\begin{equation}
 \mathcal{A}_{n}^{RL} (\mathbf{k}) + \mathcal{A}_{n}^{LR} (\mathbf{k}) = 0,
\end{equation}
where we employed that Eq.~\eqref{eq:tplus_LR} connects the same momenta when $PT_+$ symmetry is present.
Furthermore, Eq.~\eqref{eq:tplus_RR} is modified to
\begin{equation}
 \mathcal{A}_n^{RR} (\mathbf{k}) = - \mathcal{A}_n^{RR} (\mathbf{k}) + i \frac{\partial_\mathbf{k} I_{nn} (\mathbf{k})}{I_{nn} (\mathbf{k})},
\end{equation}
which gives
\begin{equation}
 \mathcal{A}^{RR} (\mathbf{k}) -\frac{i}{2} \frac{\partial_\mathbf{k} I_{nn} (\mathbf{k})}{I_{nn} (\mathbf{k})} = 0 ,
\end{equation}
as quoted in the main text.
Thus, both terms that enter the anomalous contribution to the velocity [Eq.~\eqref{eq:rdot1D}] are zero.
As we show in Appendix~\ref{sec:gauge_invariance}, the anomalous contribution is gauge-independent, thus, although we derived it here in a certain gauge choice [Eq.~\eqref{eq:ptplus_gauge}], the final result is independent of this choice.

The anomalous contribution to the weight rate is generally nonzero.

\subsection{Inversion and time-reversal $P C_+$}

Only certain terms vanish in presence of $PC_+$ symmetry with $(PC_+)(PC_+)^* =+1$.
For unbroken $PC_+$ symmetry, left and right eigenstates are related: In particular, the inverse of $U_\mathbf{k}$ may be written as
\begin{equation}
 D_\mathbf{k} U_\mathbf{k}^{-1} = U_\mathbf{k}^{T} C_+^\dagger P^\dagger,
 \label{eq:pc_symmetry}
\end{equation}
which is an extension of Eq.~\eqref{eq:cr_plus_phase}. The matrix $D_\mathbf{k}$ is a diagonal matrix that is necessary to keep left and right eigenstates orthonormal, i.e., $D_\mathbf{k}$ is included to keep the relation $U_\mathbf{k}^{-1} U_{\mathbf{k}} = 1$ intact.
The diagonal matrix $D_\mathbf{k}$ can be absorbed by $U_\mathbf{k} \to U_\mathbf{k} \sqrt{D_\mathbf{k}} $ as long as $\sqrt{D_\mathbf{k}}$ is well-defined (its elements $[D_\mathbf{k}]_{nn}$ cannot be real and negative, which is true apart from certain isolated points in parameter space).
Absorbing $D_\mathbf{k}$ changes Eq.~\eqref{eq:pc_symmetry} to the simpler $U_\mathbf{k}^{-1} = U_\mathbf{k}^{T} C_+^\dagger P^\dagger$.
Thus, in the presence of unbroken $P C_+$ symmetry, Eq.~\eqref{eq:tplus_LR} is modified in that it now connects the same momentum $\mathbf{k}$, giving
\begin{equation}
 \mathcal{A}_{n}^{LR} (\mathbf{k}) = -\mathcal{A}_{n}^{LR} (\mathbf{k}),
\end{equation}
and accordingly $\mathcal{A}_n^{LR} (\mathbf{k}) = \mathcal{A}_{n}^{RL} (\mathbf{k})= 0$.
Furthermore, Eq.~\eqref{eq:cplus_overlap} gets promoted to
\begin{equation}
 I_{nn^\prime } (\mathbf{k}) = [I^{-1}]_{n^\prime n} (\mathbf{k}) ,
\end{equation}
such that we obtain for the remaining nonzero contribution to the velocity term, using Eq.~\eqref{eq:cplus_RR},
\begin{align}
 \frac{\partial_\mathbf{k} I_{nn} (\mathbf{k})}{I_{nn} (\mathbf{k})}
 = -i \left( \mathcal{A}_{n}^{RR} (\mathbf{k}) + \mathcal{A}_n^{LL} (\mathbf{k}) \right) , \\
 \mathcal{A}^{RR}_n - \frac{i}{2} \frac{\partial_\mathbf{k} I_{nn} (\mathbf{k})}{I_{nn} (\mathbf{k})}
 = \frac{1}{2} \left( \mathcal{A}^{RR}_n - \mathcal{A}^{LL}_n \right) .
\end{align}
In fact, the eigenfunctions in the SSH model considered in Sec.~\ref{sec:numerics} can be chosen such that $\mathcal{A}_{n}^{LR} (\mathbf{k}) = \mathcal{A}_{n}^{RL} (\mathbf{k}) =0$.
Note that this is a gauge choice: The diagonal matrix $D_\mathbf{k}$ introduced above does not need to be absorbed by $U_\mathbf{k}$, which gives generally nonzero Berry connections $\mathcal{A}_n^{RL}$ and $\mathcal{A}_n^{LR}$, but leaves the gauge-independent anomalous weight rate and velocity invariant.

\subsection{Pseudo-Hermiticity}

When pseudo-Hermiticity [Eq.~\eqref{eq:pseudo_hermiticity}] is present, the left and right eigenvectors are related to one another by a unitary transformation.
Taking the Hermitian conjugate of Eq.~\eqref{eq:diagonalization_ham} gives
\begin{align}
 \diag ( \{ \varepsilon_{n,\mathbf{k}}^* \} )
 &= U_\mathbf{k}^\dagger H_\mathbf{k}^\dagger (U_\mathbf{k}^\dagger)^{-1} \\
 &= U_\mathbf{k}^\dagger \eta^\dagger H_\mathbf{k} (U_\mathbf{k}^\dagger \eta^\dagger)^{-1} ,
\end{align}
thus, for each eigenstate with energy $\varepsilon_{n,\mathbf{k}}$, there must another eigenstate with energy $\varepsilon_{n^\prime ,\mathbf{k}}=\varepsilon_{n,\mathbf{k}}^*$.
Similar to $PT_+$ and $PC_+$ symmetry, we call pseudo-Hermiticity \emph{unbroken} when these states are the same.
This allows us to fix the gauge via
\begin{equation}
D_\mathbf{k} U_\mathbf{k}^{-1} = U_\mathbf{k}^\dagger \eta^\dagger,
\label{eq:pseudo_hermitian_gauge}
\end{equation}
with a diagonal matrix $D_\mathbf{k}$, which is only possible if $\eta=\eta^\dagger$ as also required, e.g., in Ref.~\onlinecite{Kawabata:2019en}.
This implies that $D_\mathbf{k}$ is real since $D_\mathbf{k}=U_\mathbf{k}^\dagger \eta^\dagger U_\mathbf{k}$.
Since the entries $[D_\mathbf{k}]_{nn}$ are real, $\sqrt{D_\mathbf{k}}$ is in general not well defined: Differently from the $PC_+$ symmetry discussed above, the scenario ill-defined square root for $[D_\mathbf{k}]_{nn}<0$ is no longer confined to isolated points in parameter space. Hence, we cannot absorb the matrix $D_\mathbf{k}$ in the eigenvectors $U_\mathbf{k}$.
Using Eq.~\eqref{eq:pseudo_hermitian_gauge}, we can rewrite the Gramian matrix
\begin{align}
 I_{nn^\prime } (\mathbf{k})
 &= \sum_j [U_\mathbf{k}^*]_{jn} [U_\mathbf{k}]_{jn^\prime } 
  = \sum_j [ D_\mathbf{k} U_\mathbf{k}^{-1}]_{nj}  [ D_\mathbf{k} U_\mathbf{k}^{-1}]^*_{n^\prime j} \nonumber \\
 &= [D_\mathbf{k}]_{nn} [I^{-1}]_{nn^\prime } (\mathbf{k}) [D_\mathbf{k}]_{n^\prime n^\prime  },
\end{align}
thus, $I (\mathbf{k}) D_{\mathbf{k}}^{-1} = D_\mathbf{k} I^{-1} (\mathbf{k})$ and therefore squares to the identity.
We further obtain that the Berry connection transforms as
\begin{align}
 \mathcal{A}_n^{LR} (\mathbf{k})
 &= i\sum_j [ U_\mathbf{k}^{-1}]_{nj} \partial_\mathbf{k} [U_\mathbf{k}]_{jn} \nonumber \\
 &= i\sum_j [U_\mathbf{k} D_\mathbf{k}^{-1}]^*_{jn} \partial_\mathbf{k} [ D_\mathbf{k} U_\mathbf{k}^{-1}]_{nj}^* \nonumber \\
 &=  \mathcal{A}_n^{RL} (\mathbf{k}) + i [D_\mathbf{k}^{-1}]_{nn} \partial_\mathbf{k} [D_\mathbf{k}]_{nn} ,
\end{align}
and
\begin{align}
 \mathcal{A}_n^{RR} (\mathbf{k})
 &= i \frac{\sum_j [ U_\mathbf{k}^*]_{jn} \partial_\mathbf{k} [U_\mathbf{k}]_{jn}}{I_{nn} (\mathbf{k})} \nonumber \\
 &= i \frac{\sum_j [ D_\mathbf{k} U_\mathbf{k}^{-1}]_{nj} \partial_\mathbf{k} [ D_\mathbf{k} U_\mathbf{k}^{-1}]^*_{nj} }{[D_\mathbf{k} I^{-1} (\mathbf{k}) D_\mathbf{k}]_{nn}} \nonumber \\
 &=  \mathcal{A}_n^{LL} (\mathbf{k}) + i [D_\mathbf{k}^{-1}]_{nn} \partial_\mathbf{k} [D_\mathbf{k}]_{nn} .
\end{align}
The anomalous weight rate and velocity are therefore generally nonzero.
When the matrix $D_\mathbf{k}$ can be absorbed into the eigenstates, i.e., when $[D_\mathbf{k}]_{nn}>0$, the anomalous weight rate must be zero; in this situation, $I (\mathbf{k})$ equals the identity and $\mathcal{A}^{RL}=\mathcal{A}^{LR}$ as well as $\mathcal{A}^{RR} = \mathcal{A}^{LL}$.

\bibliography{references}
\end{document}